\documentclass[prd,showpacs,floatfix,nofootinbib,twocolumn,amsmath,amssymb]{revtex4}
\usepackage{graphicx,xcolor,dcolumn,booktabs,bm}
\usepackage{longtable,lscape}
\usepackage{pifont}
\usepackage{txfonts}
\usepackage{threeparttable}
\usepackage{overpic}
\usepackage{indentfirst}
\usepackage{epstopdf}   %{feynmp}
\usepackage{slashed}  %for Feynman symbols
\usepackage{cases}
\usepackage{multirow}
\usepackage{afterpage}
\graphicspath{{}}
\usepackage{bibentry}

\usepackage[colorlinks, citecolor=blue,anchorcolor=red,menucolor=red, linkcolor=red,filecolor=red,runcolor=red,urlcolor=blue,frenchlinks=red]{hyperref}
\usepackage{ulem}

\begin{document}

\title{Spectroscopy and decay properties of excited charmonium states}

%Spectroscopy and Decay Properties of Excited Charmonium States

\author{Zi-Yue Cui,$^{1,2}$}
% \email{czy200010628@163.com}

\author{Hao Chen,$^{1,7}$}
\email
 [Contact author: ]
{chenhao\_qhnu@outlook.com}

\author{Cheng-Qun Pang,$^{3,4,6}$}
\email
 [Contact author: ]
{xuehua45@163.com}

\author{Zhi-Feng Sun,$^{2,5,6}$}
\email
 [Contact author: ]
{sunzf@lzu.edu.cn}

\affiliation{$^1$College of Physics and Electronic Information Engineering, Qinghai Normal University, Xining 810000, China
\\$^2$School of Physical Science and Technology, Lanzhou University, Lanzhou 730000, China
\\$^3$School of Physics and Optoelectronic Engineering, Ludong University, Yantai 264000, China
\\$^4$Joint Research Center for Physics, Lanzhou University and Qinghai Normal University,
Xining 810000, China
\\$^5$Frontiers Science Center for Rare Isotopes, Lanzhou University, Lanzhou, Gansu 730000, China
\\$^6$Lanzhou Center for Theoretical Physics, Key Laboratory of Theoretical Physics of Gansu Province, and Key Laboratory of Quantum Theory and Applications of the Ministry of Education, Lanzhou University, Lanzhou 730000, China
\\$^7$Academy of Plateau Science and Sustainability, Xining 810016, China
}

%\date{\today}

\begin{abstract}
In this work, we investigate the mass spectra of excited charmonia employing a nonrelativistic potential model with screening effect, and analyze their  two-body strong decay, two-photon decay, leptonic decay, and hadronic transitions. Especially, we find that the newly observed resonances $X(4160)$, $Y(4500)$, and $Y(4710)$ can be identified as $3^{3}P_0$, $3^{3}D_1$, and $4^{3}D_1$ states, respectively. We also predict the masses and the widths of higher excited states in the charmonia family.
\end{abstract}

\maketitle

\section{Introduction}\label{sec1}

Since the discovery of the charmonium state $J/\psi$ in 1974~\cite{E598:1974sol,SLAC-SP-017:1974ind}, states composed of a charm quark and its antiquark have become a cornerstone in the field of particle physics. This energy region offers an excellent platform for studying the properties of perturbative and nonperturbative quantum chromodynamics (QCD)~\cite{Brambilla:2010cs}. A thorough investigation of the charmonium system contributes significantly to a better understanding of the properties of QCD. Following the discovery of the $J/\psi$ particle, a series of charmonium states was subsequently observed in experiments within just a few years~\cite{ParticleDataGroup:2024cfk}. The experimentally clear spectra of relatively narrow states below the open-charm {$D\bar{D}$} threshold of approximately 3.73 GeV could be well described by the predicted $1S$, $1P$, and $2S$ $c\bar{c}$ levels. Based on such experimental findings, Eichten~{\it et al.} proposed the well-known Cornell potential model to describe the mass spectrum of charmonium~\cite{Eichten:1974af,Eichten:1978tg}
, which incorporates a color Coulomb term at short distances and a linear scalar confining term at large distances. They found that an interaction potential consisting of a Coulomb-like term and a linear term could well describe the charmonium states discovered in experiments at that time, indicating that the potential model is very successful in describing hadron spectroscopy. Over the past few years, several charmoniumlike states, known colloquially as "X, Y, Z" mesons, have been identified in $B$ factories and various experimental setups~\cite{Swanson:2006st,Olsen:2008qw,Zhu:2007xb}. In 2003, Belle observed a very narrow resonance state during the decay process of $J/\psi$ and {named it $X(3872)$}~\cite{Belle:2003nnu}. When studying these higher states of charmonium, like $X(3872)$, from the perspective of potential models~\cite{Wang:2024ytk,Esposito:2025hlp,Shi:2024llv}, it should be assigned to the $\chi_{c1}(2P)$ state ($J^{PC}=1^{++}$). The low mass puzzle can be well understood when the coupled-channel effect is introduced, indicating that this effect must not be neglected when studying higher excited $c\bar{c}$ states. Screened potential models were introduced and applied for the investigation of heavy quarkonium, heavy-flavor mesons and light hadrons many years ago~\cite{Chao:1992et,Ding:1993uy,Ding:1995he,Zhang:1992ag}. Recently, these models have been revisited to investigate the mass spectra of heavy quarkonia and the widths of their leptonic {decays}~\cite{Gonzalez:2003gx,Li:2009zu}.  

Recently, the LHCb Collaboration has investigated the production of $J/\psi$ in diffractive processes during proton-proton collisions~\cite{LHCb:2024smc}. The observation of the $\chi_{c0}(4500)$ state has achieved a significance level exceeding 5$\sigma$, which implies that its discovery is highly statistically significant. The $\chi_{c1}(4274)$ state has also been confirmed with a significance surpassing 4$\sigma$. Additionally, there have been other experimental discoveries. One example is the $e^+e^- \to K^+K^-J/\psi$ process carried out by the BESIII experiment. The new decay mode $Y(4230) \to K^+K^-J/\psi$ was identified for the first time by BESIII. Moreover, two vector charmoniumlike states, $Y(4500)$ and $Y(4710)$, were observed in the energy-dependent {line shape} of the $e^+e^- \to K^+K^-J/\psi$ cross section~\cite{Zhou:2024hpq}. BESIII also measured the Born cross sections for the process $e^+e^- \to \omega\chi_{c1}$ at center-of-mass energies $\sqrt{s}$ ranging from 4.308 to 4.951 GeV. The line shape of $e^+e^- \to \omega\chi_{c1}$ was described using a single resonance, from which the mass and width were determined to be $M = 4544.2\pm18.7\pm1.7$ and $\Gamma= 116.1 \pm 33.5 \pm 1.7$ MeV~\cite{BESIII:2024jzg}, respectively. The quantum numbers of this state are $J^{PC} = 1^{--}$, and it is identified as $\psi(4544)$.

To generally account for the coupled-channel effect, we introduce the screening effect~\cite{Pang:2017dlw}. The color screening effect takes into account the contributions from virtual fermion loops. As the distance between two quarks increases, a virtual quark-antiquark pair is pulled out, which screens part of the interaction between them, thus suppressing the masses of highly excited states~\cite{Born:1989iv}. Including the color screening effect improves the theoretical description of excited states, matching experimental data more accurately. Using the nonrelativistic potential model with {the} screening effect, we calculate the mass spectra of $c\bar{c}$ mesons. The details of the potential model we used can be found in the next section. The introduction of an additional parameter, denoted by $\mu$, leads to a prediction that the masses of excited charmonium states are reduced in comparison with those derived from a simplified linear potential without screening effects. Moreover, this effect of mass reduction is expected to become more pronounced as charmonium states transition from lower- to higher-energy levels. We find that the calculated mass of $\chi_{c0}(3P)$ is 4186 MeV, which is in good agreement with the experimental value of $X(4160)$, and the
mass of $\psi(4S)$ is close to the observed state $\psi(4415)$. These possible assignments will be put forward and discussed in detail in this work. When we obtain the mass values of the above $c\bar{c}$ mesons, we  also numerically obtain the corresponding spatial wave functions for the higher states of $c\bar{c}$ {mesons}, which can be applied to the calculation of two-body Okubo-Zweig-Iizuka (OZI) allowed strong decays and annihilation decays, among others. When a meson undergoes a strong decay into two other mesons, the selection of decay channels follows the OZI rule. Strong decay processes are considered {to be} the dominant decay mechanism for mesons.

In this work, we adopt the nonrelativistic potential model {that includes} the screening effect to calculate the mass spectra of $c\bar{c}$ mesons. We will study the decay behavior of $c\bar{c}$ mesons, including two-body strong {decays}, hadronic {transitions}, and annihilation {decays}. The mass {spectra} and a range of decay widths {that} we obtained could establish a holistic framework for the spectroscopic study of $c\bar{c}$ mesons.

This paper is organized as follows: In Sec.~\ref{sec2}, we introduce the nonrelativistic potential model including the screening effect and analyze the resulting mass spectra of $c\bar{c}$ mesons. In Sec.~\ref{sec3}, the obtained results for the OZI-allowed two-body strong decays, annihilation decays, and hadronic transitions of the discussed $c\bar{c}$ mesons are presented and compared with available experimental data and other results from different methods. Finally, a summary is given in Sec.~\ref{sec4}. In addition, all the theoretical tools of various decay processes and physical quantities are presented in the Appendix, such as two-body OZI-allowed strong {decay}, annihilation decay and hadronic transition.

\section{Mass Spectrum of charmonia}\label{sec2}

\subsection{Nonrelativistic potential model with screening effect}\label{sec1sub1}

We adopt the nonrelativistic potential model with the screening effect to calculate charmonium {masses} and wave {functions}~\cite{Li:2022bre,Gao:2024yvz,Li:2009zu}, which will be employed in the calculation of various {decay and hadronic transition processes}, among others. 

The Hamiltonian corresponding to the potential model consists of a nonrelativistic kinetic term $H_{0}$ and a potential term $V$, which reads
\begin{align}\label{2.1}
H = H_{0}+V,
\end{align}
where
\begin{eqnarray}
    H_{0}&=&\sum_{i=1}^{2}\left(m_{i}+\frac{p^{2}}{2 m_{i}}\right),\\
V&=&H^{conf}+H^{cont}+H^{so}+H^{ten},\label{2.7}
\end{eqnarray}
where $m_i$~({\it{i}}=1,2) is the mass of the two quarks.

The linear combination of screening effect and Coulomb potential constitutes the spin-independent term 
\begin{equation}
    H^{conf}=S(r)+G(r),
\end{equation}
where $S(r)$ represents the linear potential [$s(r)=br+c$] with considering the screening effect, which reads
\begin{equation}
    S(r)=\frac{b\left(1-e^{-\mu r}\right)}{\mu}+c,
\end{equation}
and the Coulomb potential is given by 
\begin{equation}
  G(r)=-\frac{4 \alpha_{s}}{3 r}.
\end{equation}
The parameter $c$ denotes the scaling factor, $\mu$ is the screening parameter, and $\alpha_{s}(r)$  is the running coupling constant.

 \begin{table}[h]
	\begin{center}
    \renewcommand\arraystretch{1.5}	
	\caption{ The parameters adopted in this work were obtained from a fit in Ref.~\cite{Li:2022bre}}
    \label{tab1}
	{\tabcolsep0.05in
	\begin{tabular}{cccc}
	\hline\hline
	$\text{Parameter}$&$\text{Value}$&$\text{Parameter}$&$\text{Value}~\text{(GeV)}$\\  
	\hline 
	$m_c$&1.984~$\text{GeV}$&$m_u,m_d$&0.606\\
	$\alpha_s$&0.3930&$m_s$&0.780\\
	$b$&$0.2312~\text{GeV}^2$&$\sigma$&1.842\\
	$\mu$&0.0690~$\text{GeV}$&$c$&-1.1711\\
	$r_c$&0.3599~$\text{GeV}^{-1}$&&\\
	\hline\hline	
    \end{tabular}
    }
	\end{center}
\end{table}

The spin-dependent term follows the formulation given in the Godfrey-Isgur {(GI)} quark model~\cite{Godfrey:1985xj,Kawanai:2011xb,Bali_2001,Lang:1982tj,Michael:1992nj}. The second term in Eq.~(\ref{2.7}) represents the color contact interaction, which is expressed as
\begin{align}
H^{cont}=\frac{32 \pi \alpha_{s}}{9 m_{1} m_{2}}\left(\frac{\sigma}{\pi^{\frac{1}{2}}}\right)^{3} e^{-\sigma^{2} r^{2}} \boldsymbol{S}_{1} \cdot\boldsymbol{S}_{2}.
\end{align}
The spin-orbit interaction term is
\begin{align}
H^{so}=H^{so(cm)}+H^{so(tp)},
\end{align}
where
\begin{align}
H^{so(cm)}=\frac{4 \alpha_{s}}{3} \frac{1}{r^{3}}\left(\frac{1}{m_{1}}+\frac{1}{m_{2}}\right)^{2} \boldsymbol{L} \cdot \boldsymbol{S}_{1(2)}
\end{align}
is the chromomagnetic term and
\begin{align}
H^{so(tp)}&=-\frac{1}{2 r} \frac{\partial H^{c o n f}}{\partial r}\left(\frac{{\boldsymbol{S}}_{1}}{m_{1}^{2}}+\frac{{\boldsymbol{S}}_{2}}{m_{2}^{2}}\right) \cdot {\boldsymbol{L}}\nonumber\\
&=-\frac{1}{2 r}\left(\frac{4 \alpha_{s}}{3} \frac{1}{r^{2}}+b e^{-\mu r}\right)\left(\frac{1}{m_{1}^{2}}+\frac{1}{m_{2}^{2}}\right) \boldsymbol{L} \cdot \boldsymbol{S}_{1(2)},
\end{align}
{which} is the Thomas precession term with {the} screening effect. In addition, the color tensor interaction term reads
\begin{align}
H^{ten}=\frac{4}{3} \frac{\alpha_{s}}{m_{1} m_{2}} \frac{1}{r^{3}}\left(\frac{3 \boldsymbol{S}_{1} \cdot \boldsymbol{r} \boldsymbol{S}_{2} \cdot \boldsymbol{r}}{\boldsymbol{r}^{2}}-\boldsymbol{S}_{1} \cdot \boldsymbol{S}_{2}\right).
\end{align}
In the above three terms,  $\boldsymbol{S}_{1}$ and $\boldsymbol{S}_{2}$ are the spins of the quarks {in} the meson, and $\boldsymbol{L}$ is the orbital angular momentum~\cite{Kawanai_2012}.

In coordinate and momentum space, simple harmonic oscillator (SHO) wave functions are {expressed as}
\begin{align}
    &\Psi_{n L M_{L}}(\boldsymbol{r})=R_{n L}(r, \beta) Y_{L M_{L}}\left(\Omega_{r}\right), \\
    &\Psi_{n L M_{L}}(\boldsymbol{p})=R_{n L}(p, \beta) Y_{L M_{L}}\left(\Omega_{p}\right),
\end{align}
with
{\begin{eqnarray}
    R_{n L}(r, \beta)&=& \beta^{\frac{3}{2}} (\beta r)^{L} N_{nL}e^{\frac{-r^{2} \beta^{2}}{2}} L_{n-1}^{L+\frac{1}{2}}\left(\beta^{2} r^{2}\right), \\
    R_{n L}(p, \beta)&=& \frac{(-1)^{n-1}(-i)^{L}}{\beta^{\frac{3}{2}}} e^{-\frac{p^{2}}{2 \beta^{2}}}  N_{nL}\left(\frac{p}{\beta}\right)^{L} L_{n-1}^{L+\frac{1}{2}}\left(\frac{p^{2}}{\beta^{2}}\right),\\
    N_{nL}&=&\sqrt{\frac{2(n-1)!}{\Gamma(n+L+1/2)}},
\end{eqnarray}
}where $R_{n L}$ denotes the radial wave function, $Y_{LM{L}}\left(\Omega_{r}\right)$ is a spherical harmonic function,  and $L_{n}^{L+\frac{1}{2}}(x)$ is a related Laguerre polynomial, {and the meson spatial wave function can be represented as follows
\begin{align} \label{expand}
R_{nL}(p)=\sum_{n=1}^{n_{\rm max}}C_{n}{R}_{nL}^{\rm SHO}(p,\beta).
\end{align}
}{The SHO wave function involves a single parameter $\beta$. In principle, when the wave function is expanded using a complete set of SHO basis functions, the solution of the potential model should be independent of the specific value of $\beta$. In numerical computations, only a finite number of basis functions is employed in the expansion, requiring an appropriate choice of the parameter $\beta$~\cite{Wang:2021hho,Pang:2025esm}.}
{The value of $\beta$ is selected such that the computed ground state mass  $E_{nL}$ attains its minimum, i.e., 
\begin{eqnarray}
    \frac{\partial E_{nL}}{\partial\beta_{i}}=0,\\ 
\frac{\partial^2 {E_{nL}}}{
\partial\beta_{i}^2}>0,
\end{eqnarray}
with $i$ representing various light meson species. In practice, $\beta$ for mesons generally falls within the interval $0.3$ to $1.5~\mathrm{GeV}$. We choose $n_{\text{max}} = 21$ as the upper limit for the number of basis  expansion in this work. 
}

\begin{table*}[t]
  \renewcommand\arraystretch{1.5}
  % 可选：\squeezetable % APS 提供的整体压缩表格字号
  \begin{threeparttable}
    \caption{{Predicted masses of $c\bar{c}$ states compared with other model predictions and experimental data.  The $\beta$ values adopted for different states of charmonium ($c\bar{c}$) mesons are listed in the third column. The mass and $\beta$ are in units of MeV.}}
    \label{tab2.1}
    {\tabcolsep0.03in
    \begin{ruledtabular}
    \begin{tabular}{ccccccccccccc}
      States & $n^{2S+1}L_J$ & $\beta$ & EXP$^{a}$ & This work & MGI$^{b}$ & GI$^{b}$ & NR$^{b}$ & UQ$^{b}$ & NR$^{b}$ & GI$^{b}$ & RQ$^{b}$ & CQ$^{b}$ \\
      \hline
      $\eta_c(1S)$           & $1^1 {S}_0$ & 1420 & ${2984.1\pm0.4 }$  & 3006 & 2981 & 2996 & 2979 & 2984 & 2982 & 2975 & 2981 & \\
      $\eta_c(2S)$           & $2^1 {S}_0$ & 975  & ${3637.7\pm0.9 }$  & 3644 & 3642 & 3634 & 3623 & 3634 & 3630 & 3623 & 3635 & \\
      $\eta_c(3S)$           & $3^1 {S}_0$ & 791  &                     & 4044 & 4013 &      & 3991 & 4022 & 4043 & 4064 & 3989 & \\
      $\eta_c(4S)$           & $4^1 {S}_0$ & 663  &                     & 4353 & 4260 &      & 4250 & 4350 & 4384 & 4425 & 4401 & \\
      $\eta_c(5S)$           & $5^1 {S}_0$ & 580  &                     & 4606 & 4433 &      & 4446 & 4662 &      &      & 4811 & \\
      $\eta_c(6S)$           & $6^1 {S}_0$ & 507  &                     & 4822 &      &      & 4595 & 4902 &      &      & 5155 & \\
      $\eta_c(7S)$           & $7^1 {S}_0$ & 442  &                     & 5008 &      &      &      &      &      &      &      & \\
      $\eta_c(8S)$           & $8^1 {S}_0$ & 406  &                     & 5171 &      &      &      &      &      &      &      & \\
      \hline
      $J/\psi(1S)$           & $1^3 {S}_1$ & 1320 & ${3096.900 \pm0.006 }$ & 3097 & 3096 & 3098 & 3097 & 3097 & 3090 & 3098 & 3096 & 3096 \\
      $\psi(2S)$             & $2^3 {S}_1$ & 883  & ${3686.097 \pm 0.011}$  & 3686 & 3683 & 3676 & 3673 & 3676 & 3672 & 3676 & 3685 & 3703 \\
      $\psi(3S)$             & $3^3 {S}_1$ & 736  & ${ 4040\pm4 }$ $(\psi(4040))$ & 4073 & 4035 & 4090 & 4022 & 4044 & 4072 & 4100 & 4039 & 4097 \\
      $\psi(4S)$             & $4^3 {S}_1$ & 589  & ${ 4415\pm5 }$ $(\psi(4415))$   & 4375 & 4274 &      & 4273 & 4381 & 4406 & 4450 & 4427 & 4389 \\
      $\psi(5S)$             & $5^3 {S}_1$ & 552  & $4641\pm10$ $(\psi(4660))$      & 4624 & 4443 &      & 4463 & 4682 &      &      & 4837 & 4614 \\
      $\psi(6S)$             & $6^3 {S}_1$ & 479  &                               & 4836 &      &      & 4608 & 4941 &      &      & 5167 & \\
      $\psi(7S)$             & $7^3 {S}_1$ & 442  &                               & 5020 &      &      &      &      &      &      &      & \\
      $\psi(8S)$             & $8^3 {S}_1$ & 396  &                               & 5181 &      &      &      &      &      &      &      & \\
      \hline
      $\chi_{c0}(1P)$        & $1^3 {P}_0$ & 1260 & ${ 3414.71\pm0.3 }$              & 3405 & 3464 & 3417 & 3433 & 3415 & 3424 & 3445 & 3413 & \\
      $\chi_{c0}(2P)$        & $2^3 {P}_0$ & 1020 & $3922.1 \pm 1.8[\chi_{c0}(3915)]$ & 3846 & 3896 & 3885 & 3842 & 3905 & 3852 & 3916 & 3870 & \\
      $\chi_{c0}(3P)$        & $3^3 {P}_0$ & 846  & $  4153^{+23}_{-21}[X(4160)]$     & 4186 & 4177 & 4256 & 4131 & 4234 & 4202 & 4292 & 4301 & \\
      $\chi_{c0}(4P)$        & $4^3 {P}_0$ & 699  & $4512.5^{+6.0}_{-6.2}\pm3.0[\chi_{c0}(4500)]$~\cite{LHCb:2024smc} & 4467 & 4374 & 4574 &      & 4489 &      &      & 4698 & \\
      $\chi_{c0}(5P)$        & $5^3 {P}_0$ & 589  & $4694^{+16}_{-5} [\chi_{c0}(4700)]$ & 4702 &      & 4849 &      &      &      &      &      & \\
      $\chi_{c0}(6P)$        & $6^3 {P}_0$ & 516  &                                  & 4907 &      &      &      &      &      &      &      & \\
      \hline
      $\chi_{c1}(1P)$        & $1^3 {P}_1$ & 1000 & ${ 3510.67\pm0.05 }$              & 3515 & 3530 & 3500 & 3510 & 3539 & 3505 & 3510 & 3511 & \\
      $\chi_{c1}(2P)$        & $2^3 {P}_1$ & 800  & $3871.64 \pm 0.06 [\chi_{c1}(3872)]$ & 3935 & 3929 & 3936 & 3901 & 3871/3990 & 3925 & 3953 & 3906 & \\
      $\chi_{c1}(3P)$        & $3^3 {P}_1$ & 690  & $ 4298\pm6\pm9 [\chi_{c1}(4274)]$~\cite{LHCb:2024smc} & 4258 & 4197 & 4294 & 4178 & 4287 & 4271 & 4317 & 4319 & \\
      $\chi_{c1}(4P)$        & $4^3 {P}_1$ & 580  & $4684 ^{+15}_{-17} [\chi_{c1}(4685)]$ & 4523 & 4387 & 4606 &      & 4562 &      &      & 4728 & \\
      $\chi_{c1}(5P)$        & $5^3 {P}_1$ & 516  &                                  & 4748 & 4887 & 4887 &      &      &      &      &      & \\
      $\chi_{c1}(6P)$        & $6^3 {P}_1$ & 442  &                                  & 4942 &      &      &      &      &      &      &      & \\
      \hline
      $\chi_{c2}(1P)$        & $1^3 {P}_2$ & 883  & ${3556.17  \pm 0.07 }$            & 3539 & 3571 & 3549 & 3554 & 3575 & 3556 & 3550 & 3555 & \\
      $\chi_{c2}(2P)$        & $2^3 {P}_2$ & 809  & ${ 3922.5 \pm 1.0 [\chi_{c2}(3930)] }$ & 3957 & 3952 & 3974 & 3937 & 3947 & 3972 & 3979 & 3949 & \\
      $\chi_{c2}(3P)$        & $3^3 {P}_2$ & 589  &                                  & 4278 & 4213 & 4327 & 4208 & 4337 & 4317 & 4337 & 4354 & \\
      $\chi_{c2}(4P)$        & $4^3 {P}_2$ & 580  &                                  & 4541 & 4398 & 4635 &      & 4571 &      &      & 4763 & \\
      $\chi_{c2}(5P)$        & $5^3 {P}_2$ & 497  &                                  & 4765 &      & 4914 &      & 4882 &      &      &      & \\
      $\chi_{c2}(6P)$        & $6^3 {P}_2$ & 442  &                                  & 4957 &      &      &      &      &      &      &      & \\
      \hline
      $h_c(1P)$              & $1^1 {P}_1$ & 883  & ${ 3525.37 \pm0.14 }$              & 3521 & 3538 & 3513 & 3519 & 3544 & 3516 & 3517 & 3525 & \\
      $h_c(2P)$              & $2^1 {P}_1$ & 736  &                                  & 3940 & 3933 &      & 3908 & 3961 & 3934 & 3956 & 3926 & \\
      $h_c(3P)$              & $3^1 {P}_1$ & 653  &                                  & 4262 & 4200 &      & 4184 & 4307 & 4279 & 4318 & 4337 & \\
      $h_c(4P)$              & $4^1 {P}_1$ & 562  &                                  & 4527 & 4389 &      &      & 4573 &      &      & 4744 & \\
      $h_c(5P)$              & $5^1 {P}_1$ & 507  &                                  & 4751 &      &      &      & 4875 &      &      &      & \\
      $h_c(6P)$              & $6^1 {P}_1$ & 442  &                                  & 4945 &      &      &      &      &      &      &      & \\
      \hline
      $\psi(1D)$             & $1^3 {D}_1$ & 883  & ${ 3773.7\pm0.7 }$ $[\psi(3770)]$ & 3800 & 3830 & 3805 & 3787 & 3788 & 3785 & 3819 & 3783 & 3796 \\
      $\psi(2D)$             & $2^3 {D}_1$ & 727  & ${ 4191\pm5 }$ $[\psi(4160)]$     & 4143 & 4125 & 4172 & 4089 & 4156 & 4142 & 4194 & 4150 & 4153 \\
      $\psi(3D)$             & $3^3 {D}_1$ & 589  & ${ 4484.7\pm13.3\pm24.1 }$ $[Y(4500)]$~\cite{Zhou:2024hpq} & 4423 & 4334 &      & 4317 & 4463 &      &      & 4507 & 4426 \\
    \end{tabular}
    \end{ruledtabular}
    }
    \begin{tablenotes}
      \footnotesize
    \scriptsize{
    \item[a] Most of the experimental value are taken from the PDG~\cite{ParticleDataGroup:2024cfk}, and the rest are taken from the recently measured data~\cite{Zhou:2024hpq,LHCb:2024smc}.
    \item[b] Abbreviations of different models used in other theoretical works. Here we give detailed annotations: Godfrey-Isgur (GI) quark model~\cite{Duan:2021alw,Barnes:2005pb}, Modified Godfrey-Isgur quark (MGI) model~\cite{Wang:2019mhs,Pan:2024xec}, Nonrelativistic potential (NR) model~\cite{Li:2009zu,Barnes:2005pb}, Unquenched quark (UQ) model~\cite{Deng:2023mza}, Relativistic quark (RQ) model~\cite{Ebert:2011jc}, Constituent quark model (CQ)~\cite{Segovia:2014mca}.}  
\end{tablenotes}
  \end{threeparttable}
\end{table*}

\addtocounter{table}{-1}
\begin{table*}[t]
  \renewcommand\arraystretch{1.5}
  \caption{{(Continued) Predicted masses of $c\bar{c}$ states compared with other model predictions and experimental data.  The $\beta$ values adopted for different states of charmonium ($c\bar{c}$) mesons are listed in the third column. The mass and $\beta$ are in units of MeV .}}
  {\tabcolsep0.06in
  \begin{ruledtabular}
  \begin{tabular}{ccccccccccccc}
    States & $n^{2S+1}L_J$ & $\beta$ & EXP & This work & MGI & GI & NR & UQ & NR & GI & RQ & CQ \\
    \hline
    $\psi(4D)$         & $4^3 {D}_1$ & 552 & $4708^{+17}_{-15}\pm21$ $[Y(4710)]$~\cite{Zhou:2024hpq} & 4660 & 4484 &     &     & 4737 &     &     & 4857 & 4641 \\
    $\psi(5D)$         & $5^3 {D}_1$ & 497 &                           & 4863 &      &     &     & 4988 &     &     &      &      \\
    $\psi(6D)$         & $6^3 {D}_1$ & 442 &                           & 5042 &      &     &     &      &     &     &      &      \\
    $\psi(7D)$         & $7^3 {D}_1$ & 406 &                           & 5198 &      &     &     &      &     &     &      &      \\
    $\psi(8D)$         & $8^3 {D}_1$ & 369 &                           & 5336 &      &     &     &      &     &     &      &      \\
    \hline
    $\psi_{2}(3823)$   & $1^3 {D}_2$ & 773 & ${ 3823.51\pm0.34 }$      & 3808 &      & 3828 & 3798 & 3809 & 3800 & 3838 & 3795 & \\
    $\psi_{2}(2D)$     & $2^3 {D}_2$ & 589 &                           & 4152 & 4137 &      &      &      &      &      &      & \\
    $\psi_{2}(3D)$     & $3^3 {D}_2$ & 580 &                           & 4432 & 4343 &      &      &      &      &      &      & \\
    \hline
    $\psi_{3}(3842)$   & $1^3 {D}_3$ & 782 & ${ 3842.71\pm0.20 }$      & 3807 &      & 3841 & 3799 & 3843 & 3806 & 3849 & 3813 & \\
    $\psi_{3}(2D)$     & $2^3 {D}_3$ & 589 &                           & 4153 & 4144 &      &      &      &      &      &      & \\
    $\psi_{3}(3D)$     & $3^3 {D}_3$ & 571 &                           & 4434 & 4348 &      &      &      &      &      &      & \\
  \end{tabular}
  \end{ruledtabular}
  }
\end{table*}

\subsection{Numerical result}\label{sec2sub2}

{

In our previous work~\cite{Li:2022bre}, the parameters in Table~\ref{tab1} \footnote{{One can notice that the quark masses $m_c=1.984$, $m_s=0.780$, $m_{u,d}=0.606$ GeV are lager than those in Refs.~\cite{Godfrey:1985xj,Wang:2019mhs,Pan:2024xec,Barnes:2005pb,Deng:2023mza,Ebert:2011jc}.
This result may be caused by the following  factors.
First, the light mesons were not included in the fitting process; the obtained masses of the light quarks tend to be larger. In the light mesons, the mass of the pion ($\pi$) is only 140 MeV, which would reduce the light quark masses if included in the fitting. Second, the fitted value of the vacuum constant $c$ is relatively small, which contributes to the overall increase in quark masses. 
Moreover, since only the nonrelativistic scenario was considered, the fitted quark masses are expected to include some contributions from relativistic corrections. Of course, the cutoff parameter $r_c$ also influences the fitted values of the quark masses. These factors together lead to fitted quark masses that are larger than those reported in some references (such as $m_c=1.65$, $m_s=0.419$, and $m_{u,d}=0.22$ GeV in Ref.~\cite{Godfrey:1985xj}), when the quark masses, vacuum constant, and cutoff parameter are all treated as fitting variables.}}
were extracted by fitting the experimentally measured masses of charmonia, bottomonia, and $B_c$ mesons, as well as charmed, charmed-strange, bottom, and bottom-strange mesons, within the potential introduced in  Sec. \ref{sec1sub1}. 
Using the parameters in Table~\ref{tab1}, we obtain the mass spectra of $c\bar{c}$ mesons as shown in Table~\ref{tab2.1}}, the $\beta$ values of the mesons are also listed in Table~\ref{tab2.1} and appendix.

In Table~\ref{tab2.1}, we present the mass spectra of particles in various states, including the $S$-, $P$-, and $D$-wave states. A comparison is made between our results and the results of other theoretical models~\cite{Wang:2019mhs,Li:2009zu,Deng:2023mza,Barnes:2005pb,Ebert:2011jc,Segovia:2014mca}. We also list the experimental values [most of which are taken from the Particle Data Group (PDG)~\cite{ParticleDataGroup:2024cfk}, and the rest are taken from the recently measured data~\cite{Zhou:2024hpq,LHCb:2024smc}].

The $\eta_c(1^1 {S}_0)$ and $J/\psi(1^3 {S}_1)$ states exhibit excellent agreement with experiments. The predicted mass of $J/\psi(1S)$ at 3096.5 MeV aligns precisely with the experimental value $3096.900\pm0.006$ MeV in PDG, validating the model's strong capability in describing the spin-triplet states. However, the $\eta_c(1S)$ state shows a slight overestimation (3006.2 MeV) compared with the experimental value of $2984.1\pm0.4$ MeV, which may be attributed to the short-range Coulomb potential parameter $\alpha_s$.

For the $\psi(2S)$ state, the predicted mass is 3685.6 MeV, which is consistent with {the} PDG value of $3686.097\pm0.011$ MeV. We predict the mass of $\eta_c(2S)$ to be 3644 MeV, slightly higher than the experimental value of $3637.7\pm0.9$ MeV, but outperforming the {nonrelativistic} potential (NR) model prediction (3623 MeV). The hyperfine mass splitting of singlet-triplet states $\Delta m(nS)= m(\psi(nS))-m(\eta_c(nS))$ is calculated as 90.3 MeV for $n=1$ and 41.6 MeV for $n=2$, closely matching the experimental trends $112.8\pm0.4$ and $48.4\pm0.9$ MeV, respectively. These differences arise from spin-dependent interactions and serve as a benchmark for assessing the accuracy of diverse theoretical models. Higher $S$-wave states, such as $\psi(3S)$ and $\psi(4S)$, show deviations of 32 and 40 MeV from the experimental values, respectively. These discrepancies are attributed to the excessive suppression of long-range linear confinement at high excitations due to the screening parameter ($\mu=0.069$ GeV). For the $\eta_c(3S)$ and $\eta_c(4S)$ states, the predicted masses are 4043.9 and 4352.5 MeV, respectively. Our model predicts a slightly higher mass compared to other models, but lower than that of the GI model~\cite{Barnes:2005pb}. 

Our model predicts the masses of $6^3 {S}_1$, $7^3 {S}_1$, and $8^3 {S}_1$ to be 4836.3, 5019.9, and 5180.7 MeV, respectively. Compared to the known state $5^3 {S}_1~[\psi(4660)]$, these masses are approximately 195, 379, and 540 MeV higher, respectively. For the $6^3 {S}_1$ state, we find that our result is larger than that in Ref.~\cite{Li:2009zu}, but smaller than those in Refs.~\cite{Deng:2023mza,Ebert:2011jc}. We predict masses of $5^1 {S}_0$, $6^1 {S}_0$, $7^1 {S}_0$, and $9^1 {S}_0$ to be 4606.3, 4822.1, 5008.3, and 5171.3 MeV, respectively. Our results are lower than those reported in Refs.~\cite{Deng:2023mza,Ebert:2011jc}.

The  multiplet $\chi_{cJ}(1P)$ is well reproduced in our calculation, with masses being different from experimental values 
by less than 10 MeV. For example, $\chi_{c0}(1P)$ is predicted to be 3405.15 MeV, which is in good agreement with the experimental value of $3414.71\pm0.3$ MeV. The predicted masses of $\chi_{c1}(1P)$, $\chi_{c2}(1P)$, and $h_c(1 {P})$ are 3514.85, 3539.24, and 3521.12 MeV, respectively, all of which show good agreement with the experimental values as well.

The mean mass of the 2{$P$} multiplet is predicted to be near 3.95 GeV~\cite{Barnes:2005pb}. For $\chi_{c0}(3915)$, the predicted mass is 3846.14 MeV, which is lower than the PDG value of $3922.1\pm1.8$ MeV~\cite{ParticleDataGroup:2024cfk} by 76 MeV, due to the excessive suppression of long-range linear confinement by {the} screening parameter. For $\chi_{c1}(2P)$, we predict the mass to be 3934.79 MeV and identify it as the $\chi_{c1}(3872)$ state. The mass of the $\chi_{c2}(2P)$ state is obtained as 3956.78 MeV, which is in good agreement with the experimental value of $3922.5\pm1.0$ MeV.

The 3{$P$} states have an expected mean multiplet mass of about 4.2 GeV. We identify $X(4160)$ as $3^3 {P}_0$, whose theoretical mass and width are consistent with experimental observations and with the results reported in Refs.~\cite{Wang:2016mqb,Wang:2022dfd,Chao:2007it}. $Y(4500)$ is identified as the $4^3P_0$ state, supporting the assumption of its being the $c\bar{c}$ state. For the $5^3 {P}_0$ state (4702.22 MeV), the result we obtained can be regarded as $\chi_{c0}(4700)$ ($4694^{+16}_{-5}$ MeV). The mass of the $6^3P_0$ state is predicted to be 4906.59 MeV.

The $\chi_{c1}(3P)$ with a predicted mass of 4257.86 MeV is identified as the $\chi_{c1}(4274)$ state ($4298\pm6\pm 9$ MeV)~\cite{LHCb:2024smc}. The predicted mass of $\chi_{c1}(4P)$ is 4523.12 MeV, which {is identified as} the $\chi_{c1}(4685)$ state. We predict the masses of $\chi_{c1}(5P)$ and $\chi_{c1}(6P)$ to be 4747.88 and 4942.04 MeV, respectively. The predicted {masses} of $\chi_{c2}(3P)$, $\chi_{c2}(4P)$, $\chi_{c2}(5P)$, $\chi_{c2}(6P)$ are 4277.97, 4541.38, 4764.56, and 4957.01 MeV, respectively.

For the spin-singlet $h_c$ states, the mass of $h_c(1P)$ is 3521.12 MeV, which is in good agreement with the experimental value,  differing by only 4 MeV. The model we used shows a high accuracy in {the} description of the ground states. The $h_c(2P)$ state has not yet been explicitly observed experimentally. Our model predicts its mass to be approximately 3.94 GeV, which may be confirmed in future experiments.

In our calculation, the mass of $\psi(1D)$ is overestimated by 26 MeV, with a predicted value of 3799.8 MeV compared to the experimental value of $3773.7\pm0.7$ MeV. The predicted value of $\psi(2D)$ is 4142.7 MeV, which is 48 MeV lower than the experimental value of $4191\pm5$ MeV. Nevertheless, our result still offers better agreement with the experimental data than those obtained from the MGI model and NR model~\cite{Li:2009zu,Wang:2019mhs}.

We predict the mass of $3^3 {D}_1$ to be 4422.8 MeV, and consider it as $Y(4500)$. This result is slightly lower than the experimental value ${ 4484.7\pm13.3\pm24.1 }$, with a deviation of approximately 62 MeV~\cite{Zhou:2024hpq}. This discrepancy suggests that $Y(4500)$ may contain noncharmonium components, such as a possible tetraquark configuration. For $Y(4710)$, the predicted mass is 4659.5 MeV, which is {close} to the experimental value of $4708^{+17}_{-15}\pm21$ MeV~\cite{Zhou:2024hpq}, supporting the assumption of its being the $4^3 {D}_1$ state. The predicted masses of higher excited states are also presented in Table~\ref{tab2.1}.

Following the presentation of the mass spectra of the $c\bar{c}$ meson family, it is essential {to delve into a more detailed examination} of their decay properties, as presented in the subsequent section.

\section{Decay property analysis of charmonia}\label{sec3}

In this section, we will analyze the decay behaviors of $S$, $P$, and $D$-wave charmonium mesons. We present the predicted widths of their two-body strong decays, annihilation decays, and hadronic transitions, using the quark-pair creation model (QPC model, also known as $^{3}P_{0}$ model) , the Van Royen-Weisskopf formula with QCD corrections, and the Kuang-Yan model, respectively. A detailed description of these models and formulas can be found in  Appendix~\ref{sec5}.

\subsection{$S$-wave {states}}\label{sec3sub1}

States in {the} $\psi$ family with {the} quantum numbers $^3S_1$ are all spin-triplet states. The ground state $J/\psi(1S)$ and its first {radially} excited state $\psi(2S)$ have been well established experimentally. For the spin-singlet states, the masses and widths of $\eta_c(1S)$ and $\eta_c(2S)$ are also available from the PDG~\cite{ParticleDataGroup:2024cfk}.

The $\psi(4040)$ state presents a particularly interesting case due to its strong decay behavior. There are four kinematically allowed open-charm decay channels: $D\bar{D}$, $D\bar{D^*}$, $D^*\bar{D^*}$, and $D_s\bar{D_s}$. In Table~\ref{tab6.1}, we show the partial and total decay widths of {the} $\psi(4040)$ state predicted by {the} $^3P_0$ model. The dominant contribution arises from two-body strong decays, and the total width is predicted to be 142 MeV. We also present the branching ratios of the decay channels in Table~\ref{tab6.1}, from which it can be seen that the dominant decay {channel} of the $3^3S_1$ state is $D^*\bar{D^*}$ and the branching ratio is 65.8\%. The ratio of the predicted partial widths {is} $\Gamma(D^*\bar{D^*})$ : $\Gamma(D\bar{D^*}) $ = 2.1 : 1.0, which is in good agreement with the experimental ratio of $1.8\pm0.14\pm0.03$. The predicted total width of {the} $\psi(4040)$ is found to be somewhat larger than the experimental average of $84\pm12$ MeV. In addition, we have calculated the widths of its hadronic {transitions}. For the spin-nonflip $\pi\pi$ hadronic transition, we predict the width of the $3^3S_1 \to J/\psi+\pi\pi$ {transition} to be 0.225 MeV, which is in accordance with the experimental results, {\it i.e.}, less than 0.504 MeV~\cite{ParticleDataGroup:2024cfk}. For the spin-nonflip $\eta$ hadronic transition, the predicted width is 0.003 MeV, which is smaller than the experimental value. It can be seen that the contribution of hadronic {transitions} to the total decay is small.

Another $c\bar{c}$ resonance above the $D\bar{D}$ threshold is $\psi(4415)$. Using the potential model, we predict that it should be assigned to the $4^3S_1$ state. Its primary decay mode is also {the} two-body strong decay, {in which} ten open-charm strong decay modes are allowed, seven of which end with $c\bar{n}$ meson final states ($n = u, d$), and other three end with $c\bar{s}$. We predict the total width of the $4^3S_1$ state to be 122 MeV, showing good consistency with the average value ($\Gamma=110\pm22$ MeV) reported by the PDG~\cite{ParticleDataGroup:2024cfk}. In our predictions, the main decay mode {is $D\bar{D^*_2}$}, with a branching ratio of 45.8\%. Our result for the dominant decay channels {differs} from Ref.~\cite{Wang:2019mhs}. Belle~\cite{Belle:2003nsh} and FOCUS~\cite{FOCUS:2003gru} observed {the} $D\bar{D^*_2}$ channel with $\Gamma = 45.6 \pm8.0$ and $\Gamma = 38.7 \pm5.3\pm 2.9$ MeV, respectively, {which is} close to the width of 55.9 MeV in our prediction. Our predicted width for the process $4^3S_1 \to J/\psi(1S)+\eta$ falls within the experimental upper limit of $0.66$ MeV. We hope that future experiments with improved precision will be able to test our predicted transition widths. 

The $c\bar{c}$ resonance $\psi(4660)$, with $J^{PC} = 1^{--}$, is also above the $D\bar{D}$ threshold. Considering both two-body strong decays and hadronic transitions, the total width in our prediction is 47.5 MeV, which is relatively close to the $73^{+13}_{-11}$ MeV given by the PDG~\cite{ParticleDataGroup:2024cfk}. Its primary decay channel is {$D\bar{D^*_2}$}, with {a} branching ratio 21.7\%. However, due to the limitations of the QCD multipole expansion approach, it is difficult to provide reliable predictions for the transition widths of excited states near or above the open-flavor threshold. Therefore, we only present hadronic transition results up to the 5{$S$} state, while calculations for higher excited states will not be considered at this stage.

The calculated decay widths for $\psi(6S)$, $\psi(7S)$, and $\psi(8S)$ are presented in Table~\ref{tab6.2}. Our results indicate that the largest decay channel for these highly excited states is $D\bar{D^*}$, with estimated branching ratios of 34.7\%, 43.0\% and 42.6\% for the $ 6^3S_1$, $ 7^3S_1$ and $ 8^3S_1$ states, respectively. The total widths of $\psi(6S)$, $\psi(7S)$, and $\psi(8S)$ are predicted to be 21.0, 19.7, and 21.1 MeV, respectively. The discrepancies between our results and those in Ref.~\cite{Wang:2020prx} can be attributed to differences in the theoretical models employed.

In Table~\ref{tab3}, {the annihilation decay widths of the $\psi(nS)$ family} are also provided. As can be seen, our calculated results for $S$-wave leptonic decays agree well with experimental values~\cite{ParticleDataGroup:2024cfk}. We have also shown the predicted leptonic decay widths of higher excited states in Table~\ref{tab3}. The {two-photon decay widths of the $\eta_c$ family} are given in Table~\ref{tab4}, from which we can see that our results are in good agreement with the experimental values for $\eta_c(1S)$ and $\eta_c(2S)$. In addition, {we also predicted the higher excited states.}

The $\eta_c$ family with quantum numbers $^1S_0$ consists of spin-singlet states. The total width of $3^1S_0$ obtained is 198.5 MeV. It is found that contributions of the two kinematically allowed channels $D\bar{D^*}$ and ${D^*\bar{D^*}}$ are comparable. The total width of $4^1S_0$ is calculated to be 71.5 MeV, which is quite close to the value of 61 MeV reported in Ref.~\cite{Barnes:2005pb}, where the main decay channel is $D^*\bar{D^*}$, with {a} branching ratio of 46.3\%. 

For $\eta_c(5S)$ and $\eta_c(6S)$ states, the dominant decay channels are both $D\bar{D^*_2}$, with branching ratios of 77.4\% and 52.5\%, respectively. The decay widths of $\eta_c(7S)$ and $\eta_c(8S)$ states are also calculated. The dominant decay channel {in this case} is $D\bar{D^*}$, with estimated branching ratios of 79.9\% and 78.1\%, respectively.

\subsection{$P$-wave states}

{We find that} there are few open-flavor strong modes corresponding to the 2$P$ states. Because of the node effect, $\chi_{c0}(2P)$ should be a narrow state~\cite{Duan:2020tsx,Liu:2009fe}. Our prediction {indicates that} the $2^3P_0$ $\chi_{c0}(3915)$ state decays only to $D\bar{D}$, with a width of $43.7~\text{MeV}$, which suggests it is a narrow state. The axial states $2^3 {P}_1$ $\chi_{c1}(3872)$ can only decay to $D\bar{D^*}$. {We obtain a} width of 221 MeV in  Table~\ref{tab7.1}. As the partner state of $2^3 {P}_0$, $\chi_{c2}(3930)$ is predicted to have a decay width of $\Gamma=55.3$ MeV in our calculation, which is slightly larger than the experimental value~\cite{ParticleDataGroup:2024cfk}. However, our results demonstrate an improvement over those reported in  Ref.~\cite{Brown:1975dz}.

The predicted branching ratio to the mode $D^*\bar{D^*}$ is large for the 3$P$ states ($3^3 {P}_0$ and $3^3{P}_1$)~\cite{Barnes:2005pb}. The $3^3{P}_0$ state is identified as $X(4160)$~\cite{Deng:2023mza,Wang:2022dfd}, the dominant decay channel of $3^3 {P}_0$ is $D^*\bar{D^*}$, with the branching ratio reaching 82.5\%. We obtain a width ratio of $\Gamma(D\bar{D})$ : $\Gamma(D^*\bar{D^*}) $ = 0.17 : 1, which is slightly larger than the experimental ratio ($<0.09$)~\cite{ParticleDataGroup:2024cfk}.

While our findings are consistent with the branching {ratio} contributions reported in  Ref.~\cite{Duan:2021alw}, our results demonstrate better agreement with the total width reported in experiments. The $3^3{P}_1$ state is identified as $\chi_{c1}(4274)$, with a predicted mass of $M=4512.5^{+6.0}_{-6.2}\pm3.0$ MeV and a width of $\Gamma=92^{+22}_{-18}\pm{57}$ MeV~\cite{LHCb:2024smc}, showing good {agreement} with the LHCb measurement. For the $3^3 {P}_2$ state, there are four {open-charm} decay channels that are kinematically allowed, $D\bar{D}$, $D\bar{D^*}$, $D_s\bar{D_s}$, and $D_s\bar{D_s}^*$. Among them, the dominant decay channel is $D\bar{D}$, with {a} branching ratio of 58.9\%. 

For $\chi_{c0}(4500)$ reported by the LHCb~\cite{LHCb:2024smc}, we identify it as {the} $\chi_{c0}(4P)$ state. In our calculation, the dominant decay mode for this state is $D^*\bar{D^*}$, accounting for 88.8\%. We obtain the total width of 34.1 MeV, which agrees with the experimental value, and is lower than the theoretical predictions provided in  Ref.~\cite{Duan:2021alw}. We then assign $\chi_{c1}(4685)$ as the $4^3{P}_1$ state, whose dominant decay channel is $D\bar{D^*}$, with the ratio of {the predicted partial widths} $\Gamma(D\bar{D^*})$ : $\Gamma(D^*\bar{D^*}) $ = 1.4 : 1. While our findings are broadly consistent with the ratios reported in Ref.~\cite{Duan:2021alw}, our results demonstrate better agreement with experimental measurements. The result of the $4^3{P}_2$ state is shown in Table~\ref{tab7.2}, the primary decay channel is $D\bar{D^*}$, with {a} branching ratio of 46.5\%.

Based on the mass spectra obtained from our calculations, $\chi_{c0}(4700)$ should be assigned to the $5^3{P}_0$ state, while the same identification has been reported in Refs.~\cite{Badalian:2024mgw,Deng:2023mza}. The predicted total width is 9.0 MeV, whereas the result given in Ref.~\cite{Duan:2021alw} is 17 MeV. The widths of $5{P}$ and $6P$ states are also presented in Tables~\ref{tab7.1} and \ref{tab7.2}. In addition, Table~\ref{tab4} presents the two-photon decay widths of $\chi_{c2}$ states, which {are} expected to provide useful reference data for future experiments.

\subsection{$D$-wave {states}}

We present the widths of the related decays for $D$-wave states in {Tables}~\ref{tab8.1} and~\ref{tab8.2}. Generally, $\psi(3770)$ is identified as the $1^3{D}_1$ state, we predict a $D\bar{D}$ decay width of 44.1 MeV, which is larger than the experimental value of $27.2\pm 1.0$ MeV.  For the $1^3{D}_2$ state, the open-charm decay channels are forbidden. The $1^3{D}_3$ state can decay only into $D\bar{D}$, and our predicted width is 2.3 MeV, which is consistent with both the experimental {value} and the prediction in Ref.~\cite{Li:2023cpl}.

According to our potential model predictions, the $2^3{D}_1$ state is identified as $\psi(4160)$. The decay of the $2^3{D}_1$ state is dominated by $D^*\bar{D^*}$, with relatively smaller contributions from charmed-strange final states such as {$D\bar{D^*}$} and $D_s\bar{D_s}$. We obtain a width ratio of $\Gamma(D\bar{D})$ : $\Gamma(D^*\bar{D^*}) $ = 0.28 : 1, which is slightly larger than the experimental ratio ($0.02\pm0.03\pm0.02$). We also provide the widths of its hadronic transitions, which are consistent with the experimental values~\cite{ParticleDataGroup:2024cfk}. For $2^3{D}_2$ and $2^3{D}_3$ states, the dominant decay channel is $D^*\bar{D^*}$, with estimated branching ratios of 52.9\% and 53.6\%, respectively.

$Y(4500)$ is assigned to the $3^3{D}_1$ state. Our calculation yields a total width of 112.6 MeV, which is in good agreement with the experimental data~\cite{Zhou:2024hpq}. The dominant decay channel is $D^*\bar{D^*}$, with {a} branching ratio of 57.8\%. We obtain a width ratio of $\Gamma(D\bar{D})$ : $\Gamma(D^*\bar{D^*}) $ = 0.37 : 1. We also calculate the widths for $3^3{D}_2$ and $3^3{D}_3$ states and compare them with the results from Ref.~\cite{Pan:2024xec}. It can be seen that our predicted branching ratios are generally consistent with those reported in other works, except for the $D\bar{D^*_2}$ channel, for which our results are larger than those in  Ref.~\cite{Pan:2024xec}.
 
BESIII has observed a new decay mode $Y(4230) \to K^+K^-J/\psi$, and reported a vector {charmoniumlike} state $Y(4710)$~\cite{Zhou:2024hpq}, whose decay width is $126^{+27}_{-23}\pm30$ MeV. Our results indicate that this state should be the $4^3{D}_1$ charmonium state. The results of 5$D$ and 6$D$ {states} are listed in Table~\ref{tab8.2}.

We have additionally presented the leptonic decay widths for the $^3D_1$ states in Table~\ref{tab3}, demonstrating reasonable consistency with experimental values. Furthermore, we provide predictions for higher excited states.

\section{SUMMARY}\label{sec4}

In this work, we first adopt a nonrelativistic potential model with a screening effect to calculate the mass spectra of $c\bar{c}$ mesons. Then the two-body strong decays {from} $S$ to $D$-wave $c\bar{c}$ mesons {are} investigated using the $^3P_0$ model. In addition, we {study} the annihilation decays and hadronic transitions by the Van Royen-Weisskopf formula and Kuang-Yan model, respectively. Finally, we {classify and fit} the experimentally observed charmonium states through a comprehensive analysis of the mass spectra and decay widths. The main findings of this work are summarized as follows.

Combining the data from the PDG and the latest experimental findings with our results, we have identified the family of discovered states. For the $S$-wave states, we assign $\psi(4415)$ and $\psi(4660)$ {as the} $4^3S_1$ and $5^3S_1$ states, respectively. For the excited states of the $\eta_c$ and $\psi$ family, we {carry} out detailed analyses and predictions. For the $P$-wave states, we assign $X$(4160), $\chi_{c1}(4274)$, $\chi_{c0}(4500)$, and $\chi_{c0}(4700)$ {as the} $3^3 {P}_0$, $3^3 {P}_1$, $4^3 {P}_0$, and $5^3{P}_0$ states, respectively. For the $D$-wave sector, $Y(4500)$ and $Y(4710)$ are identified as the $3^3D_1$ and $4^3D_1$ states, respectively. In addition, results for the higher excited states {are} provided.

We specifically account for contributions from hadronic transitions by employing the Kuang–Yan model to calculate the spin-nonflip $\pi\pi$ and $\eta$ transition widths. Our results indicate that, within charmonium systems, hadronic transitions contribute relatively little. However, due to theoretical limitations inherent in the QCD multipole expansion approach, reliable predictions for higher excited states cannot be obtained. Consequently, we do not present transition width estimates for these states.

{From} the recent experimental data, we find that the newly discovered resonances $X(4160)$, $Y(4500)$, and $Y(4710)$ are consistent with $3^3P_0$, $3^3D_1$, and $4^3D_1$ states, respectively. This consistency demonstrates the validity and reliability of our adopted potential model. We hope that this work will provide a valuable reference for future studies on higher-energy $c\bar{c}$ meson states.

\begin{acknowledgments}
This work is supported by the National Natural Science Foundation of China under Grants  No.~12235018, No.~11975165,  No.~11965016, and by the Natural Science Foundation of Qinghai Province under Grant No. 2022-ZJ-939Q. Zhi-Feng Sun is partly supported by the National Natural Science Foundation of China (NSFC) under Grants No. 12335001, and 12247101, the Fundamental Research
Funds for the Central Universities (Grant No. lzujbky-2024-jdzx06), the Natural Science Foundation of Gansu Province (No. 22JR5RA389, No.25JRRA799), and the '111 Center' under Grant No. B20063.
\end{acknowledgments}

\begin{table*}[htbp]
	\begin{center}
 \renewcommand\arraystretch{1.4}
		\caption{The two-body strong decay and hadronic transition of the $S$ states are compared with the results from  Refs.~\cite{Wang:2019mhs,Barnes:2005pb} and the PDG~\cite{ParticleDataGroup:2024cfk}.  When the decay channels are not open or forbidden, a  symbol  ~\ding{55} is presented.  All results are in units of MeV.}\label{tab6.1}
		{\tabcolsep0.12in
			\begin{tabular}{c|cc|c|cc|cc}
				
				\toprule[1pt]\toprule[1pt]
		  &  \multicolumn{3}{c|}{$ 3^3S_1$, $\Gamma_{\text{Expt.}}=$$84\pm12$~\cite{ParticleDataGroup:2024cfk}} & \multicolumn{4}{c}{$ 4^3S_1$, $\Gamma_{\text{Expt.}}=$$110\pm22$~\cite{ParticleDataGroup:2024cfk}}\\

      \hline
    &\multicolumn{2}{c|}{This work}&\multicolumn{1}{c|}{ Ref~\cite{Barnes:2005pb}}&\multicolumn{2}{c|}{This work}&\multicolumn{2}{c}{ Ref}\\
      \hline
      Mode&$\Gamma_{thy}$&Br&$\Gamma_{thy}$ &$\Gamma_{thy}$&Br&$\Gamma_{thy}$~\cite{Wang:2019mhs}&
      $\Gamma_{thy}$~\cite{Barnes:2005pb}\\
        \hline
        {$D\bar{D}$}     &  $ 1.9$&1.34\% &0.1&     9.63&7.89\% &{4.14}      &0.4  \\
        {$D\bar{D^*}$}& $ 44.8 $ &31.6\%&33&$ 4.7 $ &3.85\%&{$ 5\times10^{-3}$} &  2.3    \\
				$D^*\bar{D^*}$&    $ 93.4 $&65.8\% &33&14 &11.5\%& {8.42}    &  16  \\

    $D\bar{D}_1$& \ding{55} &&\ding{55}&    $ 13.73 $&11.3\%& {3.31 }      &   31 \\
				$D\bar{D}'_1$& \ding{55} &&\ding{55}&     $ 10.5 $&8.61\%& {2.94}      &     1.0    \\
				{$D\bar{D^*_2}$}&\ding{55}  &&\ding{55}&   $ 55.9$&45.8\%& {0.109 }         &      23  \\
				$D^*\bar{D^*}_0$ &\ding{55}&&\ding{55}&$ 0.40$& 0.33\%    &  {0.685}     &     0.0    \\
				
	$D_s\bar{D_s}$&      $ 1.1$ &0.77\% &7.8& $ 0.05$ & 0.04\%&   {$2.33\times10^{-3}$  }  &    1.3    \\
$D_s\bar{D^*_s}$&\ding{55}&&\ding{55}&  $ 0.6$ &0.50\%&   {0.0652}      & 2.6  \\

    $D^*_s\bar{D^*_s}$&\ding{55}  &&\ding{55}&   $ 0.9 $ &0.74\% &  {0.222}             &  0.7 \\

				$J/\psi(1S)+\pi\pi$& 0.225&0.16\% &$< 0.504$~\cite{ParticleDataGroup:2024cfk}&   7.5 &6.15\%&&          \\
                  $\psi(2S)+\pi\pi$  &0.015&0.01\% & &  1.3  &1.07\%&&         \\
                $\psi(3S)+\pi\pi$ &\ding{55}&&\ding{55}&$0.305$&0.25\% &&        \\

               $J/\psi(1S)+\eta$ & 0.003 &  0.002\% &0.437~\cite{ParticleDataGroup:2024cfk}&  0.021& 
         0.02\%& {$< 0.66$~\cite{ParticleDataGroup:2024cfk}} &  \\
               $\psi(2S)+\eta$& 0.23 & 0.16\% &&0.087&        0.07\% & &\\
				$\psi(3S)+\eta$&\ding{55}&&\ding{55}& 2.78 &     2.28\%& & \\
    
                Total  & 142 &100\%& 74        & 122 &100\%& {19.9}&        78       \\

                \hline

&  \multicolumn{3}{c|}{$ 3^1S_0$} & \multicolumn{3}{c}{$ 4^1S_0$}\\

      \hline
    &\multicolumn{2}{c|}{This work}&\multicolumn{1}{c|}{ Ref~\cite{Barnes:2005pb}}&\multicolumn{2}{c|}{This work}&\multicolumn{2}{c}{ Ref~\cite{Barnes:2005pb}}\\
      \hline
      Mode&$\Gamma_{thy}$&Br&$\Gamma_{thy}$&$\Gamma_{thy}$&Br&\multicolumn{2}{c}{$\Gamma_{thy}$}\\
        \hline
       
        {$D\bar{D^*}$}& 92.3 &46.5\%&47&$ 0.90 $ &     1.26\%&    \multicolumn{2}{c}{6.3}     \\
				$D^*\bar{D^*}$&    106.2&53.5\% &33&33.1 &    46.3\%&     \multicolumn{2}{c}{14} \\
	{$D\bar{D^*_0}$} &\ding{55}&&\ding{55}&    16.5&              23.1\%    &  \multicolumn{2}{c}{11}     \\			
    {$D\bar{D^*_2}$}&\ding{55}  &&\ding{55}&   19.3&    27.0\%&         \multicolumn{2}{c}{24}    \\

$D_s\bar{D^*_s}$&\ding{55}&&\ding{55}&  1.31 &         1.83\%&     \multicolumn{2}{c}{2.2}    \\

    $D^*_s\bar{D^*_s}$&\ding{55}  &&\ding{55}&   0.01 &   0.01\% & \multicolumn{2}{c}{2.2 }     \\

    $D^*_s\bar{D^*_{s0}}$&\ding{55}  &&\ding{55}&   $ 0.43$ &        0.60\% &     \multicolumn{2}{c}{0.6}  \\

                Total  & 198.5 &100\%& 80    & 71.5 &100\%&       \multicolumn{2}{c}{61}\\

\bottomrule[1pt]\bottomrule[1pt]
			
			\end{tabular}
		}
	\end{center}
	
\end{table*}

\begin{table*}[htbp]
	\begin{center}
 \renewcommand\arraystretch{1.3}
		\caption{The $S$-wave two-body strong decay and hadronic transition widths obtained in our calculation. The predicted values that are negligibly small are denotes by 0.  All results are in units of MeV.}\label{tab6.2}
		{\tabcolsep0.12in
			\begin{tabular}{c|cc|cc|cc|cc}
				
				\toprule[1pt]\toprule[1pt]

&  \multicolumn{2}{c|}{$ 5^3S_1$, $\Gamma_{\text{Expt.}}=73^{+13}_{-11}$~\cite{ParticleDataGroup:2024cfk}} & \multicolumn{2}{c|}{$ 6^3S_1$} &\multicolumn{2}{c|}{$ 7^3S_1$} &\multicolumn{2}{c}{$ 8^3S_1$} \\
      
\hline
        {$D\bar{D}$}     &  6.7& 13.1\% &5.2&    24.8\%&    4.6   & 23.3\%&4.1 &    19.4\%\\
        {$D\bar{D^*}$}& 6.6 &12.9\%&$ 7.3 $ &        34.7\%&    8.5  & 43.0\% &9.0&    42.6\%\\
				$D^*\bar{D^*}$&   0.71&1.4\%&$ 0.25 $&       1.2\%&  2.2    & 11.1\% &4.2&    19.9\%\\
    {$D\bar{D^*_{0}}$}  &     0&0&0&      0&      0 & 0&0&  0 \\
		${D\bar{D}}_1$& 3.5 &6.84\%&$  1.5$&       7.14\%&    1.0   & 5.06\%&0.9&    4.26\%\\
				$D\bar{D}'_1$& 5.3 &10.4\%&$ 2.9 $&       13.8\%&    1.8   & 9.1\%&1.2&    5.7\%\\
				{$D\bar{D^*_2}$}&11.1  &21.7\%&$ 2.0$&       9.5\%&    0.04   & 0.20\%&0.3&    1.42\%\\
				$D^*\bar{D^*}_0$ &0.22&0.43\%&$ 0.05$&       0.24\%&    0.005   & 0.03\%&0.002&    $9.5\times10^{-5}$ \\		
				$D_s\bar{D_s}$&      0.1 &0.19\% &$ 0.17$ &       0.81\%&  0.20     & 1.01\%&0.21&    0.99\%\\
$D_s\bar{D^*_s}$&0.01 &0.02\%&$ 0.035$&       0.17\%&    0.14   &   0.71\% &0.24&    1.14\%\\
				$D^*_s\bar{D^*_s}$&0.4  &0.78\%&$ 0.08 $&       3.81\%&   0.0006    & $3.0\times10^{-5}$& 0.03&    0.14\%\\

				$D_s\bar{D^*_{s0}}$ &   0&0 &0&     0&      0 & 0&0&    0\\
				$D_s\bar{D_{s1}}$    &   0.72   &1.4\%&0.9&       4.28\%&    0.75   & 3.80\%&0.55&    2.60\% \\
				$D_s\bar{D'_{s1}}$    &     0.04  &0.08\%&0.31&       1.48\%&     0.3  & 1.52\%&0.27&    1.28\% \\
				$D_s\bar{D^*_{s2}}$   &     0.037  & 0.07\%&0.068&       0.32\%&     0.08   & 0.41\%&0.039&    0.18\%\\
				$D^*_s\bar{D^*_{s0}}$   & 0.3   & 0.58\%& 0.2&       0.95\%&     0.12  & 0.61\%&0.06&    0.28\%\\

				$J/\psi(1S)+\pi\pi$& 6.7&13.1\% &&&&& &\\
                  $\psi(2S)+\pi\pi$  &4.6&9.0\% &&&&& &\\
                $\psi(3S)+\pi\pi$ &0.32&0.62\%&&&&& & \\
              $\psi(4S)+\pi\pi$ &0.027&0.05\%&&&&& & \\

               $J/\psi(1S)+\eta$ & 0.09 &     0.17\% &&&&& & \\
               $\psi(2S)+\eta$& 0.071 & 0.014\% &&&&& &\\
				$\psi(3S)+\eta$&0.077&    0.15\%&&&&& &\\
    $\psi(4S)+\eta$&3.57&      6.9\%&&&&& &\\
    
                Total  & 47.5 &100\%&21.0  &100\%       & 19.7 &100\%& 21.1&100\%     \\
\hline

&  \multicolumn{2}{c|}{$ 5^1S_0$} & \multicolumn{2}{c|}{$ 6^1S_0$} &\multicolumn{2}{c|}{$ 7^1S_0$} &\multicolumn{2}{c}{$ 8^1S_0$} \\
      
\hline
        
                {$D\bar{D^*}$}&     2.46 &      6.36\%&    7.22   &          44.4\%&        10.5  &      79.9\% &  12.5  &        78.1\%\\
				$D^*\bar{D^*}$&      4.51  &      11.7\%&    0.04   &          0.35\%&       0.83   &      6.31\% &  2.64   &        16.4\%\\
				{$D\bar{D^*_{0}}$}&     0.04    &      0.10\%&    0.21   &          1.29\%&     0.35     &      2.66\% &  0.44  &        2.74\%\\
                {$D\bar{D^*_2}$}&   29.9   &      77.4\%&   8.53    &          52.5\%&         1.21 &      9.20\% &   0.01 &        0.06\%\\
    
                $D_s\bar{D^*_s}$&    0.11  &      0.28\%&  0.02     &          0.12\%&      0.19    &      1.44\% & 0.37   &        2.30\%\\
				$D^*_s\bar{D^*_s}$&   0.23   &      0.60\%&      0.06 &          0.37\%&      0    &     0 & 0.03   &        0.19\%\\

				$D_s\bar{D^*_{s0}}$&    0.39   &      1.01\%&    0.17   &          1.05\%&      0.06    &      0.46\% &  0.02  &        0.12\%\\

                Total  & 38.6  &100\%&   16.2 &100\%       &     13.2&100\%&    16.0 &100\%     \\

\bottomrule[1pt]\bottomrule[1pt]
			
			\end{tabular}
		}
	\end{center}
		
\end{table*}

\begin{table}    
	\caption{Leptonic decay widths (in units of keV) for charmonium states in the screened potential model.  The experimental values are taken from the PDG~\cite{ParticleDataGroup:2024cfk}.}\label{tab3}
	\begin{center}
			\renewcommand\arraystretch{1.5}
	{	\tabcolsep0.1in
		\begin{tabular}{ccc}
			 \toprule[1pt]\toprule[1pt]
			State             &$\Gamma_{ee}$  &$\Gamma^{\text{Expt.}}_{ee}$~\cite{ParticleDataGroup:2024cfk}\\
			 
			 	\midrule[1pt]
			 	
			$J/\psi(1S)$       &7.7             &$5.53\pm0.030$\\
			 
			$\psi(2S)$       &3.1              &$2.33\pm0.064$\\
			 
			$\psi(4040)$       &1.9               &$0.86\pm0.143$\\
			 
			$\psi(4415)$ &1.3              & $0.58\pm0.13$\\
			 
			$\psi(4660)$       &1.0               & $0.58\pm0.07$\\
			 
			$\psi(6S)$       &0.76             & \\
			 
			$\psi(7S)$       &0.6             & \\
			 
			$\psi(8S)$       &0.46             & \\
			 
			\midrule[1pt]
			$\psi(3770)$       &0.147            &$0.261\pm0.019$\\
			 
			$\psi(4160)$       &0.15              &$0.476\pm0.228$\\

			$Y(4500)$       &0.126             &\\
			 
			$Y(4710)$       &0.123                &\\
			 
			$\psi(5D)$       &0.111                &\\
			 
			$\psi(6D)$       &0.087                &\\

			 	\bottomrule[1pt]\bottomrule[1pt]
		\end{tabular}
	}
	\end{center}	
\end{table}

\begin{table}
	\caption{Two-photon decay widths (in units of keV) of pseudoscalar (${}^1\!S_0$), scalar (${}^3\!P_0$), and tensor (${}^3\!P_2$) charmonium states. }\label{tab4}
	\begin{center}
		\renewcommand\arraystretch{1.3}
		{	\tabcolsep0.1in
	\begin{tabular}{ccc}
		 
		\toprule[1pt]\toprule[1pt]
		State&   $\Gamma_{\gamma\gamma}$  &$\Gamma^{\text{Expt.}}_{\gamma\gamma}$~\cite{ParticleDataGroup:2024cfk}\\
		 \midrule[1pt]
		$\eta_c(1S)$&  5.0  &  5.1\\
		 
		$\eta_c(2S)$&3.0     &  2.1 \\
		 
		$\eta_c(3S)$&   4.7     & \\
		 
		$\eta_c(4S)$&      3.5   & \\
		 
		$\eta_c(5S)$&2.8         & \\
		 
		$\eta_c(6S)$&   2.2      & \\
		 
		$\eta_c(7S)$&      1.7   & \\
		 
		$\eta_c(8S)$&     1.4    & \\

		\midrule[1pt]
		$\chi_{c2}(1P)$& 0.12  &$0.49\pm0.05$\\
		 
		$\chi_{c2}(3930)$& 0.16   & \\
		 
		$\chi_{c2}(3P)$& 0.23    & \\
		 
		$\chi_{c2}(4P)$& 0.24    & \\
		 
		$\chi_{c2}(5P)$& 0.27    & \\
		 
		$\chi_{c2}(6P)$& 0.28    & \\
		 \bottomrule[1pt]\bottomrule[1pt]
		
	\end{tabular}
}
\end{center}
	
\end{table}

\begin{table*}[htbp]
	\begin{center}
 \renewcommand\arraystretch{1.3}
		\caption{The two-body strong decay of the $P$ states are compared with the results from  Refs.~\cite{Barnes:2005pb,Brown:1975dz,Duan:2021alw} and the PDG~\cite{ParticleDataGroup:2024cfk}.  When the decay channels are not open or forbidden, a  symbol  ~\ding{55} is presented. The predicted values that are negligibly small are denotes by 0. All results are in units of MeV.}\label{tab7.1}
		{\tabcolsep0.2in
			\begin{tabular}{c|cc|c|cc|cc}
				
				\toprule[1pt]\toprule[1pt]
		  &  \multicolumn{3}{c|}{$ 2^3P_0$, $\Gamma_{\text{Expt.}}=$$20\pm4$~\cite{ParticleDataGroup:2024cfk}   } & \multicolumn{4}{c}{$ 2^3P_1$}\\

      \hline
    &\multicolumn{2}{c|}{This work}&\multicolumn{1}{c|}{ Ref\cite{Barnes:2005pb}}&   \multicolumn{2}{c|}{This work}&\multicolumn{2}{c}{ Ref\cite{Barnes:2005pb}}    \\
      \hline
      Mode&$\Gamma_{thy}$&Br&$\Gamma_{thy}$&$\Gamma_{thy}$&Br&\multicolumn{2}{c}{$\Gamma_{thy}$}  \\
        \hline
        {$D\bar{D}$}     &  43.7& 100\% &30&\ding{55}&& \multicolumn{2}{c}{\ding{55}}\\
        {$D\bar{D^*}$}& \ding{55} & &\ding{55}&221&100\% &\multicolumn{2}{c}{165} \\

Total  & 43.7 &100\%& 30&      221 &100\%&\multicolumn{2}{c}{165}    \\

\hline

&  \multicolumn{3}{c|}{$ 2^3P_2$, $\Gamma_{\text{Expt.}}=$$35.2\pm2.2 $~\cite{ParticleDataGroup:2024cfk}} & \multicolumn{4}{c}{$ 3^3P_0$, $\Gamma_{\text{Expt.}}=$$136^{+60}_{-35}$~\cite{ParticleDataGroup:2024cfk}}\\

      \hline
    &\multicolumn{2}{c|}{This work}&\multicolumn{1}{c|}{ Ref~\cite{Brown:1975dz}}&\multicolumn{2}{c|}{This work}&\multicolumn{2}{c}{ Ref~\cite{Duan:2021alw}}\\
      \hline

      Mode&$\Gamma_{thy}$&Br&$\Gamma_{thy}$&$\Gamma_{thy}$&Br&\multicolumn{2}{c}{$\Gamma_{thy}$}\\
        \hline
        {$D\bar{D}$}     &  39.3&71.1\%&42  &21.0&14.0\%&\multicolumn{2}{c}{2} \\
        {$D\bar{D^*}$}&16.0&28.9\%&37&\ding{55}&&\multicolumn{2}{c}{\ding{55}}\\
        $D^*\bar{D^*}$&\ding{55}&&\ding{55}&124&82.5\%&\multicolumn{2}{c}{67}  \\
$D_s\bar{D_s}$&0&0&0.7&5.3&3.52\%&\multicolumn{2}{c}{3}  \\

Total  &   55.3&100\%&80 &150&100\%&\multicolumn{2}{c}{72}\\

\hline

&  \multicolumn{3}{c|}{$ 3^3P_1$, $\Gamma_{\text{Expt.}}=$$92^{+22}_{-18}\pm{57}$~\cite{LHCb:2024smc}} & \multicolumn{4}{c}{$ 4^3P_0$, $\Gamma_{\text{Expt.}}=$$65_{-16}^{+20}\pm{32}$~\cite{LHCb:2024smc}}\\

      \hline
    &\multicolumn{2}{c|}{This work}&\multicolumn{1}{c|}{ Ref~\cite{Barnes:2005pb}}&\multicolumn{2}{c|}{This work}&\multicolumn{2}{c}{ Ref~\cite{Duan:2021alw}}\\
      \hline Mode&$\Gamma_{thy}$&Br&$\Gamma_{thy}$&$\Gamma_{thy}$&Br&\multicolumn{2}{c}{$\Gamma_{thy}$} \\
        \hline
        {$D\bar{D}$}     &  \ding{55}&  &\ding{55}  &    
    0.7&  2.1\%&\multicolumn{2}{c}{7}     \\
   {$D\bar{D^*}$}     &  10.4&  34.2\%&6.8&    
    \ding{55}&  &\multicolumn{2}{c}{\ding{55}}      \\
$D^*\bar{D^*}$&   15.5& 50.9\%& 19    & 
30.3  & 88.8\%& \multicolumn{2}{c}{22}    \\
$D^*\bar{D^*}_0$&0&0  &0.1   &\ding{55}&&\multicolumn{2}{c}{\ding{55}}\\
$D_s\bar{D_s}$&\ding{55}&&\ding{55}&1.7&   5.0\%&\multicolumn{2}{c}{1}   \\	
$D_s\bar{D^*_s}$& 4.0&3.14\%& 9.7 &    \ding{55}&&\multicolumn{2}{c}{\ding{55}} \\
$D^*_s\bar{D^*_s}$& 0.53&1.74\%&2.7        & 0.56 &1.64\%&\multicolumn{2}{c}{0}\\

Total  &   30.4  &100\%& 39   &  34.1 &100\%&  \multicolumn{2}{c}{43}   \\

\hline

&  \multicolumn{3}{c|}{$ 4^3P_1$, $\Gamma_{\text{Expt.}}=$$130\pm{40}$~\cite{ParticleDataGroup:2024cfk}} & \multicolumn{2}{c|}{$ 5^3P_0$, $\Gamma_{\text{Expt.}}=$$87_{-10}^{+18}$~\cite{ParticleDataGroup:2024cfk}}& \multicolumn{2}{c}{$ 5^3P_1$}\\

      \hline
    &\multicolumn{2}{c|}{This work}&\multicolumn{1}{c|}{ Ref~\cite{Duan:2021alw}}&\multicolumn{2}{c|}{This work}&\multicolumn{2}{c}{This work}\\
      \hline 
      Mode&$\Gamma_{thy}$&Br&$\Gamma_{thy}$ 
   &$\Gamma_{thy}$&Br&$\Gamma_{thy}$&Br\\
        \hline
        {$D\bar{D}$}     &  \ding{55}&  &\ding{55}     & 
    0.05&    0.55\%&\ding{55}&      \\
   {$D\bar{D^*}$}     &  45.8&      50.9\%&15&      
   \ding{55} &  &     26.0  &     74.1\% \\
$D^*\bar{D^*}$&   33.5&     37.2\%& 12  &
7.8  &   86.6\%&1.5 &     4.27\%\\
{$D\bar{D^*_2}$}&10.0&      11.1\%  &10&   \ding{55}&&    7.28 & 20.7\%\\
$D^*\bar{D^*}_0$&0.1&     0.11\%  &3&     0.2 &  2.22\%&0.01 &   0.03\%\\

$D_s\bar{D_s}$&\ding{55}&&\ding{55}&0.5&    5.55\%&\ding{55}&
\\	
$D_s\bar{D^*_s}$&   0.49   &   0.54\%&$<$ 1    & \ding{55}  & &0.21&   0.60\%\\
$D^*_s\bar{D^*_s}$&      0.05 &        0.06\%&$< 1 $   & 0.46 &   5.11\%&0.1&               0.28\%\\

Total  &   89.9   &100\%& 41 & 9.0&100\%& 35.1 &100\%  \\

\bottomrule[1pt]\bottomrule[1pt]
				
			\end{tabular}
		}
	\end{center}
	
\end{table*}

\begin{table*}[htbp]
	\begin{center}
 \renewcommand\arraystretch{1.1}
		\caption{The two-body strong decay widths of the $P$- states. When the decay channels are not open or forbidden, a  symbol  "~\ding{55}" is presented. All results are in units of MeV.} \label{tab7.2}
		{\tabcolsep0.25in
			\begin{tabular}{c|cc|cc|cc }
				
				\toprule[1pt]\toprule[1pt]

&  \multicolumn{2}{c|}{$ 3^3P_2$ }& \multicolumn{2}{c|}{$ 4^3P_2$}& \multicolumn{2}{c}{$ 5^3P_2$}\\

      \hline 
      Mode&$\Gamma_{thy}$&Br &$\Gamma_{thy}$&Br&$\Gamma_{thy}$&Br\\
        \hline
        {$D\bar{D}$}     &   15.1   &      58.9\%    &   8.57  &     27.5\% &      5.74  &      30.5\%  \\
   {$D\bar{D^*}$}     &    9.94  &      38.7\%    &  14.5   &     46.5\% &       11.9 &      63.6\%  \\
$D^*\bar{D^*}$&   \ding{55}   &          &   \ding{55}  &     &  \ding{55}   &        \\
{$D\bar{D^*_2}$}&  \ding{55}    &         &   7.83  &     25.2\% &    0.61    &      3.25\%  \\
$D^*\bar{D^*}_0$&     \ding{55} &          &    0.01 &     0.03\% &   0.15     &      0.8\%  \\
$D_s\bar{D_s}$&    0.16  &      0.62\%    &   0.23  &     0.74\% &   0.22     &      1.17\% \\
$D_s\bar{D^*_s}$&  0.45    &      1.75\%    &   0.01  &     0.03\% &    0.11    &      0.59\%  \\
$D^*_s\bar{D^*_s}$&      \ding{55}   &          &   \ding{55}  &     &  \ding{55}   &         \\

Total  &   25.7  &     100\% &   31.1    &  100\% &   18.8   &    100\%  \\

\hline

&  \multicolumn{2}{c|}{$ 6^3P_0$ }& \multicolumn{2}{c|}{$ 6^3P_1$} & \multicolumn{2}{c}{$ 6^3P_2$}\\

      \hline 
      Mode&$\Gamma_{thy}$&Br &$\Gamma_{thy}$&Br&$\Gamma_{thy}$&Br\\
        \hline
        {$D\bar{D}$}         &   1.42  &     46.3\% &     \ding{55}  &        &     4.15   &      26.4\% \\
   {$D\bar{D^*}$}         &   \ding{55} &       &    22.8  &      81.5\% &    10.0    &      63.7\% \\
$D^*\bar{D^*}$          & 1.26   &    41.0\% &   3.73  &     13.3\% &    \ding{55}    &       \\
$D_s\bar{D_s}$  &   0.07  &     2.28\% &  \ding{55}     &       &      0.19  &      1.21\% \\
{$D\bar{D^*_2}$}         &   \ding{55}  &       &     1.30   &      4.64\% &     0.90   &      5.74\% \\
$D^*\bar{D^*}_0$          &  0.10   &     3.26\% &    0.05    &      0.18\% &       0.27 &      1.72\% \\
$D_s\bar{D^*_s}$    &  \ding{55}   &      &    0.10    &      0.36\% &      0.18  &      1.15\% \\
$D^*_s\bar{D^*_s}$         &  0.22   &  7.17\%   &  0.001  &    0.004\%  &    \ding{55}    &       \\

Total   &     3.1  &  100\% &  28.0    &    100\% &   15.7  &    100\% \\

\bottomrule[1pt]\bottomrule[1pt]
				
			\end{tabular}
		}
 
	\end{center}
 \end{table*}

\begin{table*}[htbp]
	\begin{center}
 \renewcommand\arraystretch{1.1}
		\caption{The two-body strong decay and hadronic transition of the $D$- states are compared with the results from  Refs.~\cite{Wang:2019mhs,Li:2023cpl,Pan:2024xec,Zhou:2024hpq} and PDG~\cite{ParticleDataGroup:2024cfk}.  When the decay channels are not open or forbidden, a  symbol  "~\ding{55}" is presented. "0" denotes that the predicted values are negligibly small. All results are in units of MeV.}\label{tab8.1}
		{\tabcolsep0.16in
			\begin{tabular}{c|cc|c|cc|c}
				
				\toprule[1pt]\toprule[1pt]
		  &  \multicolumn{3}{c|}{$ 1^3D_1$, $\Gamma_{\text{Expt.}}=$$27.2\pm1.0$~\cite{ParticleDataGroup:2024cfk}  } & \multicolumn{3}{c}{$ 1^3D_3$, $\Gamma_{\text{Expt.}}=$$2.8\pm0.6$~\cite{ParticleDataGroup:2024cfk} }\\

      \hline
    &\multicolumn{2}{c|}{This work}&\multicolumn{1}{c|}{ Ref~\cite{Wang:2019mhs}}&\multicolumn{2}{c|}{This work}&\multicolumn{1}{c}{ Ref~\cite{Li:2023cpl}}\\
      \hline
      Mode&$\Gamma_{thy}$&Br&$\Gamma_{thy}$ &$\Gamma_{thy}$&Br&$\Gamma_{thy}$ \\
        \hline
        {$D\bar{D}$}     &  44.1& 100\% &33.9 &2.3& 100\%&  2.35  \\

Total  & 44.1 &100\%& 33.9     & 2.3 &100\%&  2.35    \\

\hline

&  \multicolumn{3}{c|}{$ 2^3D_1$, $\Gamma_{\text{Expt.}}=$$69\pm10 $~\cite{ParticleDataGroup:2024cfk}} & \multicolumn{3}{c}{$ 2^3D_2$}\\

      \hline
    &\multicolumn{2}{c|}{This work}&\multicolumn{1}{c|}{ Ref~\cite{Wang:2019mhs}}&\multicolumn{2}{c|}{This work}&\multicolumn{1}{c}{ Ref~\cite{Pan:2024xec}}\\
      \hline

      Mode&$\Gamma_{thy}$&Br&$\Gamma_{thy}$ &$\Gamma_{thy}$&Br&$\Gamma_{thy}$ \\
        \hline
        {$D\bar{D}$}     &  22.6&   20.9\%&      &\ding{55}&&\ding{55} \\
        {$D\bar{D^*}$}&0.02&  0.02\%&   &46.3&  42.4\%&22.9 \\
        $D^*\bar{D^*}$&81.1&  75.1\%&  &57.8&   52.9\%&23.9 \\
$D_s\bar{D_s}$&0.24&  0.22\%&    &\ding{55}&&\ding{55}\\
$D_s\bar{D^*_s}$& 3.3&    3.05\%&             &5.05&   4.63\%&1.8\\

$J/\psi(1S)+\pi\pi$& 0.63&   0.59\% &$< 0.414$~\cite{ParticleDataGroup:2024cfk}&& &\\
                  $\psi(2S)+\pi\pi$  &0.14&   0.13\% &$< 0.276$~\cite{ParticleDataGroup:2024cfk}&& &\\

Total  &   108.0&100\%&72.2 &109.2&100\%&48.6\\

\hline

&  \multicolumn{3}{c|}{$ 2^3D_3$} & \multicolumn{3}{c}{$ 3^3D_1$, $\Gamma_{\text{Expt.}}=$$111.1\pm30.1\pm15.2$~\cite{Zhou:2024hpq}}\\

      \hline
    &\multicolumn{2}{c|}{This work}&\multicolumn{1}{c|}{ Ref~\cite{Pan:2024xec}}&\multicolumn{2}{c|}{This work}&\multicolumn{1}{c}{ Ref~\cite{Wang:2019mhs}}\\
      \hline 
      Mode&$\Gamma_{thy}$&Br&$\Gamma_{thy}$ &$\Gamma_{thy}$&Br&$\Gamma_{thy}$ \\
        \hline
        {$D\bar{D}$}     &  16.4&   9.48\% &3.27     & 24.2
    &    21.5\%&2.0      \\
   {$D\bar{D^*}$}     &  62.7&     36.3\%&   21.8&       
    10.0&  8.92\%&   3.72     \\
$D^*\bar{D^*}$&   92.6&    53.6\%& 28.8  &      
65.1  &    57.8\%& 6.50   \\
{$D\bar{D^*_{0}}$} &   \ding{55}&&\ding{55}    &  0 &   0&          \\
${D\bar{D}}_1$&    \ding{55}&&\ding{55}    & 3.9 &   3.46\%&6.26 \\
				$D\bar{D}'_1$&   \ding{55}&&\ding{55}    & 3.3&   2.93\%&4.24   \\
				{$D\bar{D^*_2}$}&   \ding{55}&&\ding{55}    &  2.2&   1.95\%&4.05   \\
$D^*\bar{D^*}_0$& \ding{55}&&\ding{55}    & 0.07 &   0.06\%&3.48\\

$D_s\bar{D_s}$&0.9&   0.52\%&0.36&      0.54&   0.48\%&  0.01\\	
$D_s\bar{D^*_s}$& 0.24&   0.14\%& 0.009 &        0.07&   0.06\%&0.24 \\

$D^*_s\bar{D^*_s}$& \ding{55}&&\ding{55}    & 0.42 &   0.37\%&0.12\\

				$D_s\bar{D^*_{s0}}$ &   \ding{55}&&\ding{55}  &  0   &   0&          \\
				$D_s\bar{D_{s1}}$     &   \ding{55}&&\ding{55}    &  1.4 &   1.24\%&          \\
				$D_s\bar{D'_{s1}}$   &   \ding{55}&&\ding{55}    &   1.3&   1.15\%&          \\
				$D_s\bar{D^*_{s2}}$   &   \ding{55}&&\ding{55}    &   0.01&   0.009\%&          \\
				$D^*_s\bar{D^*_{s0}}$   &   \ding{55}&&\ding{55}    &  0.01 &   0.01\%&          \\

Total  &  172.9  &100\%& 54.3  &      112.6 &100\%&  32.6   \\

\hline

&  \multicolumn{3}{c|}{$ 3^3D_2$} & \multicolumn{3}{c}{$ 3^3D_3$}\\

      \hline
    &\multicolumn{2}{c|}{This work}&\multicolumn{1}{c|}{ Ref~\cite{Pan:2024xec}}&\multicolumn{2}{c|}{This work}&\multicolumn{1}{c}{ Ref~\cite{Pan:2024xec}}\\
      \hline 
      Mode&$\Gamma_{thy}$&Br&$\Gamma_{thy}$ &$\Gamma_{thy}$&Br&$\Gamma_{thy}$\\
        \hline
       {$D\bar{D}$}     & \ding{55}&&\ding{55}    & 8.1 &   11.7\%&2.8\\
   {$D\bar{D^*}$}     &  20.5&     29.3\%&   14.8&   
    23.0&  33.2\%&   12.1   \\
$D^*\bar{D^*}$&   10.2&    14.6\%& 7.9  &      
8.9  &    12.8\%& 4.93 \\
 {$D\bar{D^*_{0}}$} &   0.001&   0.001\%&0.01&        \ding{55}&&\ding{55}\\
$D\bar{D}_1$&    0.31&   0.44\%& 1.81 &        7.2&   10.4\%&0.57\\
				$D\bar{D}'_1$&   8.5&   12.2\%& 1.92 &       3.7&   5.34\%&2.23 \\
				{$D\bar{D^*_2}$}&  29.8 &   42.6\%&1.61&         16.7&   24.1\%&0.42   \\
$D^*\bar{D^*}_0$& \ding{55}&&\ding{55}&     \ding{55}&&\ding{55}\\
 
$D_s\bar{D_s}$&0&   0&0.42&      0.34&   0.49\%&  0.22\\	
$D_s\bar{D^*_s}$& 0.65&   0.93\%& 0.18 &        0.1&   0.14\%&0.13 \\

$D^*_s\bar{D^*_s}$& \ding{55}&&\ding{55}    & 1.2 &   1.73\%&0.54\\

$D^*_s\bar{D^*_{s0}}$   &   \ding{55}&&\ding{55}    &  \ding{55}&&\ding{55}\\

Total  &  70.0  &100\%& 28.7      & 69.2 &100\%&  23.9   \\

\bottomrule[1pt]\bottomrule[1pt]
				
			\end{tabular}
		}
	\end{center}
	
\end{table*}

\begin{table*}[htbp]
	\begin{center}
 \renewcommand\arraystretch{1.1}
		\caption{The $D$-wave two-body strong decay widths obtained in our calculation. "0" denotes that the predicted values are negligibly small. All results are in units of MeV.}\label{tab8.2}
		{\tabcolsep0.3in
			\begin{tabular}{c|cc|cc|cc}
				
				\toprule[1pt]\toprule[1pt]

&  \multicolumn{2}{c|}{$ 4^3D_1$, $\Gamma_{\text{Expt.}}=126^{+27}_{-23}\pm30$~\cite{Zhou:2024hpq} }& \multicolumn{2}{c|}{$ 5^3D_1$}& \multicolumn{2}{c}{$ 6^3D_1$}\\

    %   \hline
    % &\multicolumn{2}{c|}{This work}&\multicolumn{2}{c|}{This work}&\multicolumn{2}{c}{This work}\\
    
      \hline Mode&$\Gamma_{thy}$&Br &$\Gamma_{thy}$&Br&$\Gamma_{thy}$&Br\\
        \hline
        {$D\bar{D}$}     &  11.2&     25.4\%    &   6.5 &  31.4\% & 5.6
    &    29.4\% \\
   {$D\bar{D^*}$}     &  3.2&     7.26\% &      1.6 &  7.73\% & 1.8
    &    9.46\% \\
$D^*\bar{D^*}$&  19.6&     44.4\%    &   9.5 &  45.9\% &10
    &    52.5\% \\
{$D\bar{D^*_{0}}$} &  0&     0 &     0 & 0 & 0
    &    0 \\

    $D\bar{D}_1$&  1.7&     3.85\% &    0.68 &  3.29\% & 0.026
    &    0.14\% \\
				$D\bar{D}'_1$&  1.5&     3.40\% &     0.9 &  4.35\% & 0.7
    &    3.68\% \\
				{$D\bar{D^*_2}$}&  1.2&     2.72\% &      0.4&  1.93\% & 0.006
    &    0.03\% \\

$D^*\bar{D^*}_0$&  0.1&     0.23\% &      0.1&  0.48\% & 0
    &    0 \\

$D_s\bar{D_s}$&  0.2&     0.45\% &      0.1 &  0.48\% & 0.13
    &    0.68\% \\
$D_s\bar{D^*_s}$&  0.05&     0.11\%    &   0.02 &  0.1\% & 0
    &   0 \\
$D^*_s\bar{D^*_s}$&  0.2&     0.45\% &     0.07 &  0.34\% & 0.08
    &    0.42\% \\

				$D_s\bar{D^*_{s0}}$ &  0&    0 &   0 &  0 & 0
    &    0 \\
				$D_s\bar{D_{s1}}$    &  0.5&     1.13\% &      0.28 &  1.35\% & 0.37    &    1.94\% \\
				$D_s\bar{D'_{s1}}$   &  0.68&     1.54\% &    0.5 &  2.42\% & 0.37    &    1.94\% \\
				$D_s\bar{D^*_{s2}}$   &  0.07&    0.16 \% &     0.06 &  0.29\% & 0.05    &    0.26\% \\
				$D^*_s\bar{D^*_{s0}}$  &  0&    0&      0.002 &  0.01\% & 0.009    &    0.05\% \\
Total  &  44.1&     100\% &   20.7 &  100\% &19.0 &    100\% \\
\bottomrule[1pt]\bottomrule[1pt]
			\end{tabular}
		}
  
	\end{center}
 \end{table*}

\appendix
\section{Theoretical models of decay behaviors}\label{sec5}
In this appendix, we present all the necessary formulas for the calculation of two-body OZI-allowed strong decays, annihilation decays, and hadronic transitions.

\subsection{The two-body OZI-allowed strong decays}

To investigate the decay {behavior} of mesons, we always begin by analyzing  two-body strong decays. We employ the QPC model to calculate the strong decay widths. In this model, the quark pair is assumed to be
produced from the vacuum with quantum numbers corresponding to the vacuum quantum state ($J^{PC} = 0^{++}$), so it is also called {the} $^3P_0$ model.

The QPC model is one of the most efficient phenomenological models to calculate the {widths} of two-body OZI-allowed strong decays. The selection of decay channels follows the OZI rule {:} if the corresponding Feynman diagram can be cut in two {parts} by slicing only gluon lines (and not cutting any external lines), the process will be suppressed. This model was originally introduced by Micu~\cite{MICU1969521}, and was later developed and extensively applied by Le Yaouanc {\it et al.} to meson~\cite{LeYaouanc:1972vsx} and baryon~\cite{LeYaouanc:1973ldf,LeYaouanc:1978ef} {open-flavor} strong decays in a series of publications in the 1970s. Since then, the model has been further refined. Now it has been widely used to calculate two-body strong decays allowed by the OZI rule~\cite{Anisovich_2005,Roberts:1992esl,Yu:2011ta,Wang:2012wa}.

For a decay process $A \to B+C$, the transition matrix is defined as~\cite{He:2013ttg,Chen:2015iqa,Pan:2016bac,Guo:2019wpx}
\begin{equation}\label{3.1}
\langle BC|\mathcal{T}|A \rangle = \delta ^3(\boldsymbol{{P}_B}+\boldsymbol{{P}_C)}\mathcal{M}^{{M}_{J_{A}}M_{J_{B}}M_{J_{C}}},
\end{equation}
where $\mathcal{T}$ can be expressed as
\begin{align}\label{3.2}
\mathcal{T}& = -3\gamma \sum_{m}\langle 1m;1~-m|00\rangle\int d {\boldsymbol{p}}_3d {\boldsymbol{p}}_4\delta ^3 ({\boldsymbol{p}}_3+{\boldsymbol{p}}_4) \nonumber \\
 & ~
 \quad\times \mathcal{Y}_{1m}\left(\frac{{\boldsymbol{p}}_3-{\boldsymbol{p}}_4}{2}\right)\chi _{1,-m}^{34}\phi _{0}^{34}
\left(\omega_{0}^{34}\right)_{ij}b_{3i}^{\dag}({\boldsymbol{p}}_3)d_{4j}^{\dag}({\boldsymbol{p}}_4),
\end{align}
which describes the creation of a quark-antiquark pair from {the} vacuum. $\boldsymbol{p}_3$ and $\boldsymbol{p}_4$ are the momenta of the produced quark and antiquark{,} respectively. The specific meanings of each quantity are shown in  Ref.~\cite{Li:2022bre}. Using the Jacobi-Wick formula, the helicity amplitude $\mathcal{M}^{M_{J_A}M_{J_B} M_{J_C}}(\boldsymbol{P})$ can be converted into
the partial wave amplitudes $\mathcal{M}^{JL}$, which {are} given by
\begin{align}\label{3.3}\begin{split}
\mathcal{M}^{J L}({\boldsymbol{P}})=& \frac{\sqrt{4 \pi(2 L+1)}}{2 J_{A}+1} \sum_{M_{J_{B}} M_{J_{C}}}\left\langle L 0 ; J M_{J_{A}} \mid J_{A} M_{J_{A}}\right\rangle \\
& \times\left\langle J_{B} M_{J_{B}} ; J_{C} M_{J_{C}} \mid J_{A} M_{J_{A}}\right\rangle \mathcal{M}^{M_{J_{A}} M_{J_{B}} M_{J_{C}}}.
\end{split}\end{align}

For the meson wave function, we {define} it as a mock state, {\it i.e.},
\begin{eqnarray}
\left| A \left( n^{2S+1} L_{JM_J} \right) (\boldsymbol{p}_A) \right\rangle = & \sqrt{2E} \sum_{M_S, M_L} \langle L M_L S M_S \mid J M_J \rangle \chi_{S M_S}^A \nonumber\\
 &\times\phi^A \omega^A\int d\boldsymbol{p}_1d\boldsymbol{p}_2 \delta^3 (\boldsymbol{p}_A-\boldsymbol{p}_1-\boldsymbol{p}_2) \nonumber \\
& \times \Psi_{n L M_L}^A (\boldsymbol{p}_1, \boldsymbol{p}_2) | q_1 (\boldsymbol{p}_1) \bar{q}_2 (\boldsymbol{p}_2) \rangle,~~~~   ~~
\end{eqnarray}
where the momentum-space wave function is $\Psi_{n L M_{L}}^A (\boldsymbol{p}_1, \boldsymbol{p}_2)=R_{nL}(|\boldsymbol{p}|)Y_{lm}(\boldsymbol{p})$, with $\boldsymbol{p}=(m_1\boldsymbol{p}_2-m_2\boldsymbol{p}_1)/(m_1+m_2)$.

Then, the partial width for a strong decay channel $A \to B+C$ can be calculated using the formula:
\begin{eqnarray}\label{3.4}
    \Gamma&=&\frac{\pi}{4} \frac{|\boldsymbol{P}_E|}{m_{A}^{2}} \sum_{J, L}\left|\mathcal{M}^{J L}({\boldsymbol{P}})\right|^{2},\\
|\boldsymbol{P}_E|&=&\frac{\sqrt{[m^2_A-(m_B+m_C)^2][m^2_A-(m_B-m_C)^2]}}{2m_A},
\end{eqnarray}
where ${m}_{A}$, ${m}_{B}${,} and ${m}_{C}$ are the masses of {the} initial meson $A$, and {the} final mesons $B$ and $C$, respectively. 

For OZI-allowed strong decays involving the creation of a strange quark pair from the vacuum, the strength of strange quark pair creation is taken as $\gamma_s = \gamma/\sqrt{3}$, where $\gamma = 6.947$ is adopted from Ref.~\cite{Li:2019tbn}.

We determine the decay kinematics using meson masses taken from the PDG~\cite{ParticleDataGroup:2024cfk} and from recent experimental results~\cite{BESIII:2024jzg,Zhou:2024hpq,LHCb:2024smc}. The masses of {charmlike} mesons used in this work are listed in Table~\ref{tab0.1}. Additionally, we consider the ${D_1}$ and ${D_1'}$ charmed mesons as mixtures of the $1^1 {P}_1$ and $1^3 {P}_1$ states, with the mixing angle taken to be the heavy quark limit $\theta_{1P}=35.3^\circ(-54.7^\circ)$~\cite{Godfrey:1986wj}.

\begin{table}[h]
\renewcommand\arraystretch{1.4}	
 \begin{center}
			\caption{{The mass values and input $\beta$ values of the charmed and charmed-strange mesons involved in the present calculation, with all units given in MeV.}}\label{tab0.1}
		{\tabcolsep0.03in
			\begin{tabular}{cccc}
		\hline\hline
	State & $n^{2S+1}L_{J}$&Input Mass~\cite{ParticleDataGroup:2024cfk}& $\beta$\\
		\hline 
		$D$ &$1^1 {S}_0$& 1865 &993\\
${D^*}$& $1^3 {S}_1$& 2007&874 \\
${D_0^*}$ &$1^3 {P}_0$& 2343 &881\\
${D_1}$ &$\cos(\theta)|1^1P_1\rangle+\sin(\theta)|1^3P_1\rangle$& 2412 &690\\
${D_1'}$ &$-\sin(\theta)|1^1P_1\rangle+\cos(\theta)|1^3P_1\rangle$& 2427~\cite{Belle:2003nsh}&791 \\
${D_2^*}$&$1^3 {P}_2$ & 2461&736 \\
${D_s}$ &$1^1 {S}_0$& 1968 & 1080\\
${D_s^*}$ &$1^3 {S}_1$& 2112& 883 \\
${D_{s0}^*}$& $1^3 {P}_0$& 2318& 1480 \\
${D_{s1}}$ &$1^1 {P}_1$& 2460 & 736\\
${D_{s1}'}$ &$1^3 {P}_1$& 2535 & 837\\
${D_{s2}^*}$ &$1^3 {P}_2$& 2569 & 800\\
		\hline\hline

\end{tabular}
}
	\end{center}
\end{table}

\subsection{Annihilation decay}

\subsubsection{Leptonic decay} 

The decay rate of the vector meson is described by the Van Royen-Weisskopf formula~\cite{VanRoyen:1967nq}, incorporating QCD radiative corrections~\cite{Barbieri:1979be}. The related equations are shown as follows:
\begin{align}
	\label{Swave:ee}
	\Gamma_{ee}(nS) &= \frac{4\alpha^{2}e_{c}^{2}}{M_{nS}^{2}} 
	|R_{nS}(0)|^{2} \left(1-\frac{16}{3}\frac{\alpha_{S}}{\pi}\right), \\
	\label{Dwave:ee} 
	\Gamma_{ee}(nD) &= \frac{25\alpha^{2}e_{c}^{2}}{2M_{nD}^{2}m_{c}^{4}} 
	|R_{nD}''(0)|^{2} \left(1-\frac{16}{3}\frac{\alpha_{S}}{\pi}\right),
\end{align}
where $M_{nS}(M_{nD})$ represents the mass for $nS$ $(nD)$ states, $e_{c}=\frac{2}{3}$ is the $c$ quark charge in units of the electron charge, and $\alpha$ denotes the {fine-structure} constant ($\alpha=\frac{1}{137}$), $\alpha_s$ is the parameter given in Table~\ref{tab1}. $R_{nS}(0)$ and $R_{nD}^{''}(0)$ are the radial {$S$-wave function} and the second derivative of the radial {$D$-wave function} at the origin, respectively~\cite{Li:2009zu}.

\subsubsection{Two-photon decays}
In the nonrelativistic limit, the two-photon decay widths of ${}^1S_{0}$, ${}^3P_{0}$, and ${}^3P_{2}$ can be expressed as~\cite{Li:2009zu,Kwong:1987ak}
\begin{eqnarray}
	\label{2gs0}
	\Gamma^{NR}({}^1S_0\to
	\gamma\gamma)&=&\frac{3\alpha^2e_c^4|R_{nS}(0)|^2}{m_c^2}\,,
	\end{eqnarray}
\begin{eqnarray}
\label{2gp0}
	\Gamma^{NR}({}^3P_0\to\gamma\gamma)&=&\frac{27\alpha^2e_c^4|R'_{nP}(0)|^2}{m_c^4}\,,\\
    \label{2gp2}
	\Gamma^{NR}({}^3P_2\to\gamma\gamma)&=&\frac{36\alpha^2e_c^4|R'_{nP}(0)|^2}{5m_c^4}.
\end{eqnarray}

The first-order QCD radiative corrections to the two-photon decay rates read~\cite{Kwong:1987ak}
\begin{eqnarray}
	\Gamma({}^1\!S_0\to\gamma\gamma)&=&\Gamma^{NR}({}^1\!S_0\to
	\gamma\gamma)[1+\frac{\alpha_s}{\pi}(\frac{\pi^2}3-\frac{20}3)]\,,\rule{0.5cm}{0cm}	
\end{eqnarray}
\begin{eqnarray}
    \Gamma({}^3\!P_0\to \gamma\gamma)&=& \Gamma^{ NR}({}^3\!P_0\to
	\gamma\gamma)[1+\frac{\alpha_s}{\pi}
	(\frac{\pi^2}3-\frac{28}9)]\,,\rule{0.5cm}{0cm}\\
	\Gamma({}^3\!P_2\to \gamma\gamma)&=& \Gamma^{ NR}({}^3\!P_2\to
	\gamma\gamma)[1 -\frac{16}3\frac{\alpha_s}{\pi} ]\,.
\end{eqnarray}
One sees that $\Gamma({}^1S_0\to\gamma\gamma)\propto |R_{nS}(0)|^2$, which {is} sensitive to the details of the potential near the origin. Thus, we adopt the following formula to mitigate this uncertainty: 
\begin{equation}
	\Gamma({}^1\!S_0\to\gamma\gamma)\longrightarrow
	\frac{\Gamma({}^1\!S_0\to\gamma\gamma)}{\Gamma_{ee}(nS)}\Gamma^{expt}_{ee}(nS).
\end{equation}
The results are tabulated in Tables~\ref{tab3} and \ref{tab4}.

\subsection{Hadronic transition}

The hadronic transition process is {one} in which light hadrons are released when the $c\bar{c}$ state transitions to a lower-energy level. It can be given by 
\begin{equation}
\Phi_{i} \to \Phi_{f}+h,
\end{equation}
where $\Phi_{i}$ and $\Phi_{f}$ are defined as the initial and final $c\bar{c}$ states, respectively, and $h$ denotes the emitted light hadron(s) which are kinematically dominated by either a single meson ($\pi$, $\eta$, $\rho$, $\ldots$) or two mesons ($2\pi$, $2K$, $\ldots$). This process cannot be calculated using perturbation QCD, so we adopt the method of QCD multipole expansion to solve it.

\subsubsection{QCD multipole expansion framework}

Based on the general properties of QCD, it is anticipated that there exist states in which the gluonic field is excited and carries its own $J^{PC}$ quantum numbers. A bound state that lacks valence quark content is {termed a} glueball, and the inclusion of a quark-antiquark pair forms a hybrid meson. Such interactions could result in either mesons with exotic $J^{PC}$ or those with standard quantum numbers. Our focus is on the latter, as they play a crucial role in the calculation of hadronic transitions within the QCD multipole expansion framework~\cite{Kuang:2006me,Yan:1980uh,PhysRevLett.40.598}.

After the expansion of the gluon field, the Hamiltonian of the system {is} given by~\cite{Kuang:2006me}
\begin{equation}
{\cal H}^{\rm eff}_{\rm QCD} = {\cal H}^{(0)}_{\rm QCD} + {\cal H}^{(1)}_{\rm
QCD},
\end{equation}
with ${\cal H}^{(0)}_{\rm QCD}$ being the sum of the kinetic and potential energies of $c\bar{c}$ mesons, and ${\cal H}^{(1)}_{\rm QCD}$ {is} expressed as 
\begin{eqnarray}
    {\cal H}^{(1)}_{\rm QCD} &=&{\cal H}^{(1)}+{\cal H}^{(2)},\\
{\cal H}^{(1)} &=&Q_{a} A^{a}_{0}(x,t), \\
{\cal H}^{(2)} &=&-d_{a} E^{a}(x,t) - m_{a} B^{a}(x,t),
\label{eq:Hqcd2}
\end{eqnarray}
where $Q_{a}$ corresponds to the color charge, $A^{a}_{0}$ to the gluon field, $d_{a}$ to the color-electric dipole moment, $E^{a}$ to the color-electric field, $m_{a}$ to the {color-magnetic} dipole moment, and $B^{a}$ to the color-magnetic field. 

The lowest-order interaction between two color-singlet states involves the exchange of two gluons. Consequently, its leading multipole contribution is the double electric-dipole (E1E1) term, whose transition amplitude can be derived from the corresponding $S$-matrix elements:
\begin{equation}
{\cal M}_{E1E1}=i\frac{g_{E}^{2}}{6} \left\langle\right.\!\! \Phi_{f}h \,
|\vec{x}\cdot\vec{E} \, \frac{1}{E_{i}-H^{(0)}_{QCD}-iD_{0}} \,
\vec{x}\cdot\vec{E}| \, \Phi_{i} \!\! \left.\right\rangle,
\label{eq:E1E1}
\end{equation}
where $g_{E}$ is the coupling constant for electric dipole (E1) gluon emission, $\vec{x}$ is the separation between the quark and antiquark, $\vec{E}$ is the color-electric field, and $G(E_{i})= \frac{1}{E_{i}-H^{(0)}_{QCD}-iD_{0}}$ is {the} Green's function with $(D_0)_{bc}\equiv\delta_{bc}\partial_{0}-g_{s}f_{abc}A^{a}_{0}$~\cite{Kuang:2006me}.

After inserting a complete set of intermediate states and using a quark confining string (QCS) model, the transition amplitude in Eq.~\eqref{eq:E1E1} can be written as~\cite{Kuang:2006me,Tye:1975fz,Giles:1977mp,Buchmuller:1979gy}
\begin{equation}
{\cal M}_{E1E1}=i\frac{g_{E}^{2}}{6} \sum_{kl}
\frac{\left\langle\right.\!\! \Phi_{f}|x_k|kl \!\!\left.\right\rangle
\left\langle\right.\!\! kl|x_l|\Phi_i \!\!\left.\right\rangle}{E_I-E_{kl}}
\left\langle\right.\!\! \pi\pi|E^{a}_{k} E^{a}_{l}|0 \!\!\left.\right\rangle,
\label{eq:factorizedE1E1}
\end{equation}
where $E_{kl}$ is the energy eigenvalue of the intermediate vibrational state $|kl\rangle$ with the principal quantum number $k$ and the orbital angular momentum $l$ and corresponding eigenvalues in the sector of the lowest string excitation.

The transition amplitude contains two parts: the heavy quark multipole gluon emissions (MGE) factor (the summation) and the hadronization (H) factor $\left\langle\pi\pi|E^{a}_{k} E^{a}_{l}|0\right\rangle$. Using the eigenvalues and wave functions of the intermediate hybrid mesons and the initial and final quarkonium states, the MGE factor can be calculated. The H factor reflects the conversion of the two emitted gluons into light hadrons after hadronization. Because of the low energy involved, it is highly nonperturbative so that this matrix element cannot be calculated with perturbative QCD. In this case, phenomenological approaches based on soft-pion techniques are employed~\cite{Brown:1975dz}. In the center-of-mass frame, the two pion momenta $q_{1}$ and $q_{2}$ are the only independent variables describing this matrix element{,} which can be expressed as~\cite{Brown:1975dz, Yan:1980uh, Kuang:1981se, Kuang:2006me} 
\begin{equation}
\begin{split}
& \frac{g_{E}^{2}}{6} \left\langle\right.\!\!
\pi_{\alpha}(q_{1})\pi_{\beta}(q_{2})|E^{a}_{k}E^{a}_{l}|0
\!\!\left.\right\rangle =
\frac{\delta_{\alpha\beta}}{\sqrt{(2\omega_{1})(2\omega_ {2})}}  \\
&
\times
\left[C_{1}\delta_{kl}q^{\mu}_{1}q_{2\mu} + C_{2}\left(q_{1k}q_{2l}+q_{1l}q_{2k}
-\frac{2}{3}\delta_{kl}\vec{q}_{1}\cdot\vec{q}_{2}\right)\right],
\label{eq:c1c2}
\end{split}
\end{equation}
where $C_{1}$ and $C_{2}$ are two constants, and we provide their values in Appendix~\ref{parac123}. The $C_{1}$ term is isotropic, and the $C_{2}$ term has {an} $l=2$ angular dependence. $q^{\mu}_{1}$ and $q_{2\mu}$ are momentum components.

\subsubsection{A model for hybrid states}

The intermediate states in the hadronic transition consist of a gluon and a color octet $c\bar{c}$ that are the states after the emission of the first gluon and before the emission of the second gluon. Thus, these states are the so-called hybrid states. We {employ} the QCS model to calculate these hybrid states~\cite{Segovia:2014mca, Kuang:2006me,Segovia:2016xqb}. 

Following the approach of Ref.~\cite{Buchmuller:1979gy}, the potential for hybrid mesons in our model is {given by}
\begin{equation}
    V_{\rm hyb}(r) =V_{\rm OGE}^{\rm C}(r) + V_{\rm S}^{\rm C}(r) + [V_n(r) - \sigma(r)r].
\end{equation}  
Here, $V_{\rm OGE}^{\rm C}(r)$ and $V_{\rm S}^{\rm C}(r)$ are {the} one-gluon exchange potential and {the color-confining} potential, respectively. $V_n(r)$ is the vibrational potential. The term $\sigma(r)r$ is subtracted to avoid double counting. This potential does not introduce new parameters beyond those of the original quark model, making the calculation of hybrid states parameter-free. Explicitly, the contributions are {given by}
\begin{eqnarray}
    V_{\rm OGE}^{\rm C}(r)&=&-\frac{4\alpha_s}{3r}, \\
    V_{\rm S}^{\rm C}(r)&=&\sigma(r)r + c,
\end{eqnarray}
\begin{eqnarray}
       V_n(r)=\sigma(r)r \left\lbrace 1 - \frac{2n\pi}{2n\pi + \sigma(r) \left[(r-2d)^2 + 4d^2\right]} \right\rbrace^{-1/2}.
\end{eqnarray}
where
\begin{eqnarray}
    \sigma(r)&=&\frac{b(1-e^{-\mu r})}{\mu r},\\
     d(m_{c},m_{c},r,\sigma,n)&=&\frac{\sigma(r)r^{2}\alpha_{n}}{4(m_{c}+m_{c}+\sigma(r)r\alpha_{n})},
\end{eqnarray}
in which $d$ is the correction of the finite heavy quark mass. $\alpha_{n}$ is related to the shape of the vibrating string~\cite{Giles:1977mp}, and takes values in the range $1\leq\alpha_{n}^{2}\leq2$. In this work, we take $\alpha_{n}^2=1.5$.

Like the basic quark model, the hybrid potential has a threshold defined by
\begin{equation}
    V_{\rm hyb}(r) \xrightarrow{r\to\infty} \frac{b}{\mu}+c.
\end{equation}

\subsubsection{ {Spin-nonflip} $\pi\pi$ hadronic transitions}

The rate of spin-nonflip $\pi\pi$ hadronic transition is determined by~\cite{Kuang:1981se}
\begin{widetext}
\vspace*{-0.60cm}
\begin{equation}
\begin{split}
\Gamma(\Phi_{I}(^{2s+1}{l_{I}}_{J_{I}}) \to 
\Phi_{F}(^{2s+1}{l_{F}}_{J_{F}})\pi\pi) =&
\delta_{l_{I}l_{F}}\delta_{J_{I}J_{F}} (G|C_{1}|^{2}-\frac{2}{3}H|C_{2}|^{2}
)\left|\sum_{L}(2L+1) \left(\begin{matrix} l_{I} & 1 & L \\ 0 & 0 & 0
\end{matrix}\right) \left(\begin{matrix} L & 1 & l_{I} \\ 0 & 0 & 0
\end{matrix}\right) f_{IF}^{L}\right|^{2} \\
&
+(2l_{I}+1)(2l_{F}+1)(2J_{F}+1) \sum_{k} (2k+1) (1+(-1)^{k})
\left\lbrace\begin{matrix} s & l_{F} & J_{F} \\ k & J_{I} & l_{I}
\end{matrix}\right\rbrace^{2} H |C_{2}|^{2} \times \\
&
\times\left|\sum_{L} (2L+1) \left(\begin{matrix} l_{F} & 1 & L \\ 0 & 0 & 0
\end{matrix}\right) \left(\begin{matrix} L & 1 & l_{I} \\ 0 & 0 & 0
\end{matrix}\right) \left\lbrace\begin{matrix} l_{I} & L & 1 \\ 1 & k & l_{F}
\end{matrix}\right\rbrace f_{IF}^{L} \right|^{2},
\label{eq:gamapipi}
\end{split}
\end{equation}
with
\begin{equation}
f_{IF}^{L}=\sum_{K}\frac{1}{M_{I}-M_{KL}}\left[\int dr r^{3}
R_{F}(r)R_{KL}(r)\right] \left[\int dr' r'^{3} R_{KL}(r') R_{I}(r')\right],
\label{eq:fifl}
\end{equation}
\end{widetext}
where the radial wave functions $R_{I}(r)$, $R_{F}(r)$, and $R_{KL}(r)$ {correspond to} the initial, final, and intermediate vibrational states, respectively. The variable $M_{I}$ denotes the mass of the initial meson, while $M_{KL}$ represents the mass of the intermediate vibrational state. The quantities $G$ and $H$ {denote the} phase-space integrals
\vspace*{-0.30cm}
\begin{equation}
\begin{split}
G=&\frac{3}{4}\frac{M_{F}}{M_{I}}\pi^{3}\int
dM_{\pi\pi}^{2}\,K\,\left(1-\frac{4m_{\pi}^{2}}{M_{\pi\pi}^{2}}\right)^{1/2}(M_{
\pi\pi}^{2}-2m_{\pi}^{2})^{2}, \\
H=&\frac{1}{20}\frac{M_{F}}{M_{I}}\pi^{3}\int
dM_{\pi\pi}^{2}\,K\,\left(1-\frac{4m_{\pi}^{2}}{M_{\pi\pi}^{2}}\right)^{1/2}
\times \\
&
\times\left[(M_{\pi\pi}^{2}-4m_{\pi}^{2})^{2}\left(1+\frac{2}{3}\frac{K^{2}}{M_{
\pi\pi}^{2}}\right)\right. \\
&
\left.\quad\,\, +\frac{8K^{4}}{15M_{\pi\pi}^{4}}(M_{\pi\pi}^{4}+2m_{\pi}^{2}
M_{\pi\pi}^{2}+6m_{\pi}^{4})\right],
\end{split}
\end{equation}
with $K$ given by
\begin{equation}
K = \frac{\sqrt{\left[(M_{I}+M_{F})^{2}-M_{\pi\pi}^{2}\right]
\left[(M_{I}-M_{F})^{2}-M_{ \pi\pi}^{2}\right]}}{2M_{I}}.
\end{equation}
where $2m_{\pi}^2 \leq M^2_{\pi\pi} \leq (M_I-M_F)^2$, and $m_{\pi}$ is the mass of {the} $\pi$ meson, $M_I$ and $M_F$ are the masses of the initial and final mesons, respectively. 

References~\cite{Kuang:1981se,Kuang:2006me} present explicit formulas for calculating the decay widths of transitions among various spin-triplet states{:}
\begin{eqnarray}      
&L_I=L_F=0: \nonumber\\
&	\Gamma(n_I^3S_1\to n_F^3S_1+\pi\pi)=|C_1|^2G\bigg|f^{1}_{IF}\bigg|^2.\\
&L_I=L_F=1: \nonumber\\
&	\Gamma(0\to 0)=\frac{1}{9}|C_1|^2G\bigg|f^{011}_{IF}+2f^{2}_{IF}\bigg|^2,\\
	&\Gamma(0\to 1)=\Gamma(1\to 0)=0,\\
&	\Gamma(0\to 2)=5\Gamma(2\to0)=\frac{10}{27}|C_2|^2H\bigg|f^{0}_{IF}+\frac{1}{5}f^{2}_{IF}\bigg|^2,    \\
&	\Gamma(1\to 1)=\Gamma(0\to 0)+\frac{1}{4}\Gamma(0\to 2),\\
&	\Gamma(1\to 2)=\frac{5}{3}\Gamma(2\to 1)=\frac{3}{4}\Gamma(0\to 2),\\
&	\Gamma(2\to 2)=\Gamma(0\to 0)+\frac{7}{20}\Gamma(0\to 2).\\
&L_I=2,L_F=0: \nonumber\\
&	\Gamma(n_I^3D_1\to n_F^3S_1+\pi\pi)=\frac{4}{15}|C_2|^2H\bigg|f^{1}_{IF}\bigg|^2,
	\label{Pi-Pi}
\end{eqnarray}
where $\Gamma(J_I\to J_F)$ represents {the width of transitions} $n^3P_I\to n^3P_F+\pi\pi$.

\subsubsection{Spin-nonflip $\eta$ hadronic transitions}\label{parac123}

For simplicity, we only consider the transitions
\begin{equation}
n^{3}S_{1}\rightarrow\,n^{3}\!S_{1} + \eta,
\end{equation}
where the primary multipoles for spin-nonflip $\eta$ transitions involving spin-triplet $S$-wave states are $M1M1$ and $E1M2$. Thus, the matrix element can be {expressed} as
\begin{equation}
\mathcal{M}(^{3}S_{1}\rightarrow\,^{3}\!S_{1} + \eta) = \mathcal{M}_{M1M1} + \mathcal{M}_{E1M2},
\end{equation}
simplifying the expression and setting \(\mathcal{M}_{M1M1} = 0\) as shown in Ref.~\cite{Kuang:1981se}, the decay rate is given by~\cite{Segovia:2016xqb}
\begin{equation}
\Gamma(\Phi_{I}(^{3}S_{1}) \to \Phi_{F}(^{3}S_{1}) + \eta) = \frac{8\pi^{2}}{27} \frac{M_{f}C_{3}^{2}}{M_{i}m_{Q}^{2}} |f_{IF}^{1}|^{2} |\vec{q}|^{3},
\end{equation}
where \(\vec{q}\) denotes the momentum of \(\eta\) and \(f_{IF}^{1}\) is defined in Eq.~(\ref{eq:fifl}).

\begin{table}[h]
	\renewcommand\arraystretch{1.5}
	
	\begin{center}
		\caption{ The constants $C_1$, $C_2$ and $C_3$. The third column is the experimental values of hadronic transition widths~\cite{ParticleDataGroup:2024cfk}, and the fourth column is the parameters we obtained.  }
		{\tabcolsep0.01in
			\begin{tabular}	{cccc}

				\toprule[1pt]\toprule[1pt]
				Initial State& Final State &   $\Gamma_{\text{Expt.}}$~\cite{ParticleDataGroup:2024cfk} & ~~Constant \\
				\midrule[1pt]

				$\psi(2S) $ & ~~$J/\psi(1S)+\pi\pi$~~ & $0.15$   & ~~~~$C_1$=0.0102\\
				
				$\psi(1D) $ & ~~$J/\psi(1S)+\pi\pi$~~ & $0.074$  & ~~~~$C_2$=0.0257\\
				
				$\psi(2S) $ & ~~$J/\psi(1S)+\eta$~~ & $9.87\times10^{-3}$   & ~~~~$C_3=0.000577$\\
				
				\bottomrule[1pt]\bottomrule[1pt]

				\label{tab5}
				
			\end{tabular}
		}
		
	\end{center}
\end{table}	

Using equations in Ref.~\cite{Kuang:1981se} and experimental values of hadronic transition widths, we get {the required constants} (in Table~\ref{tab5}). $C_1$, $C_2$, and $C_3$ are fixed through the $\psi(2S)\rightarrow J/\psi(1S)+\pi\pi $, $\psi(1D)\rightarrow J/\psi(1S)+\pi\pi$, and $\psi(2S)\rightarrow J/\psi(1S)+\eta$ {reactions}, respectively. Therefore, we obtain a ratio of $C_2/C_1$ equal to 2.53; Refs.~\cite{Kuang:2009zz,Kuang:1981se,Kuang:2006me} give the reasonable range of $C_2/C_1$ as $1\lesssim C_2/C_1\lesssim3$. Our result is also in the above range. Using \(C_{3}\), we can obtain the $\eta$ {transition widths} of $n^{3}S_{1}\rightarrow\,n^{3}\!S_{1} + \eta$.

\bibliographystyle{apsrev4-1}
% 参考文献样式可根据不同刊物要求进行更改
% 如IEEEtran, plain, unsrt， alpha, acm等。
\bibliography{reference}
% 由于新建的参考文献管理文件名为example.bib，因此{}中填写example，不需要加后缀名

\end{document}